%% file: chandra_snapshots_write_up_aa_for_arxiv_and_A_A_13_Aug.tex
\documentclass[usenatbib]{aa}
\usepackage{natbib} 
\usepackage[utf8]{inputenc}
\usepackage{graphicx}
\usepackage{url}
\usepackage{amsmath}
\usepackage{rotating}
\usepackage{aas_macros}
\usepackage{multirow}
\usepackage{tabularx,ragged2e,booktabs} 
\usepackage{fixltx2e} 
\usepackage{wrapfig}
\usepackage{color}
\usepackage{appendix}
\usepackage{pdflscape}
\usepackage{rotating}
\usepackage{adjustbox}

\usepackage{geometry}
\newcolumntype{L}{>{\RaggedRight\arraybackslash}X} 
\graphicspath{ {/Users/cl16977/Documents/emacs_files/chandra_snapshots/images/} }

\input{newcom}

\makeatletter

\newcommand{\Rmnum}[1]{\expandafter\@slowromancap\romannumeral #1@}
\makeatother

\begin{document} 
\newcommand*{\rom}[1]{\expandafter\@slowromancap\romannumeral #1@}


   \subtitle{The XXL Survey: \Rmnum{33}. \emph{Chandra} Constraints on the AGN Contamination of $z>1$ XXL Galaxy Clusters\thanks{Based on observations obtained with
       {\em XMM-Newton}, an ESA science mission with instruments and
       contributions directly funded by ESA Member States and NASA, and \emph{Chandra}, a NASA science mission with instruments and contributions directly funded by NASA.}}

   \author{C. H. A. Logan,
          \inst{1}
          B. J. Maughan,
          \inst{1}
          M. N. Bremer,
          \inst{1}
          P. Giles,
          \inst{1}
          M. Birkinshaw,
          \inst{1}
          L. Chiappetti,
          \inst{2}
          N. Clerc,
          \inst{3}
          L. Faccioli,
          \inst{4}
          E. Koulouridis,
          \inst{4}
          F. Pacaud,
          \inst{5} 
          M. Pierre,
          \inst{4} 
          M. E. Ramos-Ceja,
          \inst{5}
          C. Vignali, 
          \inst{6,7}
          J. Willis
          \inst{8}
        }
        
   \institute{School of Physics, HH Wills Physics Laboratory, Tyndall Avenue, Bristol, BS8 1TL,
  UK
              \\
             \email{crispin.logan@bristol.ac.uk}
             \and 
             INAF, IASF Milano, via Bassini 15, I-20133 Milano, Italy
             \and
             IRAP, Universit\'e de Toulouse, CNRS, UPS, CNES, Toulouse, France
	    \and
             AIM, CEA, CNRS, Universit\'e Paris-Saclay, Universit\'e Paris Diderot, Sorbonne Paris Cit\'e, F-91191 Gif-sur-Yvette, France
             \and
             Argelander-Institut fur Astronomie, University of Bonn,
             Auf dem Hugel 71, D-53121 Bonn, Germany 
             \and
             Dipartimento di Fisica e Astronomia, Alma Mater Studiorum, Universit\`a degli Studi di Bologna, Via Gobetti 93/2, I-40129 Bologna, Italy 
             \and
             INAF -- Osservatorio di Astrofisica e Scienza dello Spazio di Bologna, Via Gobetti 93/3, I-40129 Bologna, Italy
             \and
             Department of Physics and Astronomy, University of Victoria, 3800 Finnerty Road, Victoria, BC, Canada  
                          }

   \date{\today}

 
  \abstract
   {The XMM-XXL survey has used observations from the \emph{XMM-Newton} observatory to detect clusters of galaxies over a wide range in mass and redshift. The moderate PSF (FWHM $\sim$ 6$''$ on-axis) of \emph{XMM-Newton} means that point sources within or projected onto a cluster may not be separated from the cluster emission, leading to enhanced luminosities and affecting the selection function of the cluster survey.}
   {We present the results of short \emph{Chandra} observations of 21 galaxy clusters and cluster candidates at redshifts $z>1$ detected in the XMM-XXL survey in X-rays or selected in the optical and infra-red.}
   {With the superior angular resolution of \emph{Chandra} we investigate whether there are any point sources within the cluster region that were not detected by the XMM-XXL analysis pipeline, and whether any point sources were misclassified as distant clusters.}
   {Of the 14 X-ray selected clusters, nine are free from significant point source contamination, either having no previously unresolved sources detected by \emph{Chandra}, or with less than about 10\% of the reported XXL cluster flux being resolved into point sources. Of the other five sources, one is significantly contaminated by previously unresolved AGN, and four appear to be AGN misclassified as clusters. All but one of these cases are in the subset of less secure X-ray selected cluster detections and the false positive rate is consistent with that expected from the XXL selection function modelling. We also considered a further seven optically-selected cluster candidates associated with faint XXL sources that were not classed as clusters. Of these, three were shown to be AGN by \emph{Chandra}, one is a cluster whose XXL survey flux was highly contaminated by unresolved AGN, while three appear to be uncontaminated clusters.  
By decontaminating and vetting these distant clusters, we provide a pure sample of clusters at redshift $z>1$ for deeper follow-up observations, and demonstrate the utility of using \emph{Chandra} snapshots to test for AGN in surveys with high sensitivity but poor angular resolution.} 
   {}


   \authorrunning{C. H. A. Logan et al.}
   \titlerunning{\emph{Chandra} Constraints on the AGN Contamination of $z>1$ XXL Galaxy Clusters}
   \maketitle

\section{Introduction} \label{section.intro}

Galaxy cluster surveys provide us with large, well-controlled samples of clusters that enable us to place constraints on cosmological models through tests of the growth of structure. For the tightest constraints on the cosmological parameters, we need a large look-back time, with samples that include clusters at $z>1$. These high-redshift clusters enable the study of the astrophysical processes that drive galaxy and cluster evolution over cosmic time. 

Although galaxy cluster surveys can be carried out at different wavelengths (e.g. \citealp{Rosati:1998a,Bohringer:2004a,Gladders:2005a,Eisenhardt:2008a,Rozo:2010a,Sehgal:2013a,Planck-Collaboration:2014a,Stanford:2014a}), searching for extended X-ray emission has the advantage that the cluster candidates that are identified are much less likely to suffer from projection effects than selecting clusters based on galaxy overdensities which can contain projections of galaxies along the line of sight that are not associated with virialized systems. This is because a given amount of gas dispersed in clumps and filaments will be much fainter in X-rays than the same gas confined and compressed in a single potential well, as is the case in a cluster, where this gas is termed the intra-cluster medium (ICM). This is due to bremsstrahlung emissivity (the main emission mechanism in a cluster) being proportional to the square of the density of the gas.  

X-ray surveys have proven very effective in identifying large numbers of galaxy clusters (e.g. \citealp[][hereafter XXL Paper \Rmnum{1}]{Gioia:1990a,Ebeling:1998a,Rosati:1998a,Bohringer:2004a,Pierre:2004a,Ebeling:2010a,Fassbender:2011a,Mehrtens:2012a,Willis:2013a,Pierre:2016a}) including many at redshifts $z>1$, with the most distant clusters found up to a redshift $z \approx$ 2 \citep[][hereafter XXL Paper \Rmnum{5}]{Nastasi:2011a,Santos:2011a,Willis:2013a,Mantz:2014a}.

While X-ray surveys are effective at finding clusters of
galaxies, clusters are vastly outnumbered by active galactic nuclei
(AGN), which dominate extragalactic X-ray source counts. With
sufficient angular resolution, clusters are resolved, allowing these
two classes to be separated. However, for clusters at cosmological
distances this becomes challenging because of the low surface brightness of the cluster emission and the fact that the detected emission from these distant clusters can have angular extents similar to (or smaller than) the PSF of most X-ray observatories. This can lead to AGN being misclassified
as clusters or a compact cluster being misclassified as AGN. 

It is also possible for a genuine cluster detection to be contaminated
by X-ray emission from an unresolved AGN in, or projected onto the cluster, giving rise to various issues (e.g. \citealp{Giles:2012a}). Most importantly a cluster
with AGN contamination will have its flux and temperature overestimated \citep{Branchesi:2007a}. This has implications for the use of luminosity or temperature as a mass estimator to carry
out cosmological studies (reviewed by \citealp{Allen:2011a}), or for studies of the scaling relations between cluster properties
(e.g. \citealp[][also known as XXL Paper \Rmnum{3}]{Pratt:2009a,Maughan:2012a,Giles:2016a}). Unresolved AGN in or projected onto clusters also alter the apparent surface brightness distribution of the cluster which can enhance or decrease its detection probability making it difficult to understand the selection function of cluster surveys at the level needed for cosmological studies. An additional complication is that AGN in galaxy clusters are significantly more common at higher redshift. \citet{Galametz:2009a} found that X-ray selected AGN are at
least three times more prevalent in clusters at $1 < z < 1.5$ than in clusters at $0.5 < z < 1$. This is a higher increase in AGN density than that seen in the field population of AGN \citep{Martini:2013a}. For low mass clusters ($\lta$ 3 $\times$ 10$^{14}$ M$_{\odot}$) at z $<$ 1 there is evidence that the density of X-ray selected AGN in X-ray selected clusters is consistent with the field \citep{Koulouridis:2014a}. Optically selected AGN in optically selected clusters show similar agreement between the AGN fraction in clusters and the field \citep{Marziani:2017a}, but with some indication that the AGN fraction can be higher in compact groups \citep{Martinez:2010a}.

The problem of AGN contamination of X-ray cluster surveys can be addressed statistically by using realistic models of the population of AGN in and projected onto distant clusters in the calibration of the selection function. The state-of-the-art is the use of full cosmological hydrodynamical simulations which include self-consistent modelling of cluster and AGN populations \citep{Koulouridis:2017a}. The observational data upon which to base such models are sparse, and this project was the first systematic attempt to observationally survey the AGN content of distant X-ray selected galaxy clusters. Similar work can also now be found in \citet{Biffi:2018a}. The AGN contribution to individual distant clusters has previously been studied (e.g. \citealp{Hilton:2010a}), and the cosmic evolution of AGN in clusters has been studied using IR selected clusters, including z $>$ 1 clusters \citep{Galametz:2009a}, but this is the first time that clusters detected in an X-ray survey have been looked at, so this work has particular bearing for X-ray cluster surveys.
 
Our work uses the XXL survey (XXL Paper \Rmnum{1}), which is the largest survey carried out by the \emph{XMM-Newton} satellite and covers a total area of 50 deg$^{2}$ distributed over two fields (XXL-N and XXL-S). \emph{XMM-Newton} has an on-axis half energy width (HEW) PSF of $\sim$15$''$ which degrades and becomes increasingly asymmetric as a function of distance from the aimpoint. The XXL survey's primary aim is to investigate the large-scale structure of the Universe using the distribution of galaxy clusters (and AGN) as tracers of the matter distribution. The survey has detected several hundreds of galaxy clusters out to a redshift of $z \approx 2$  \citep[365 in the most recent list,][referred to as XXL Paper \Rmnum{20} hereafter]{Adami:2018a} above an X-ray flux limit of $\sim$ $5\times10^{-15}\flux$ in the 0.5 - 2 keV band. We study a set of 21 $z>1$ clusters and candidates using short \emph{Chandra} observations to assess the level of AGN contamination. We use the term ``candidates'' in recognition of the fact that some of the sources without spectroscopic confirmation or flagged as less reliable by the X-ray detection pipeline may not be genuine clusters. The main aims of this work are to quantify the contribution of unresolved point sources to the XXL detection of extended ICM emission and flag for rejection those candidate clusters where the XXL detection is fully resolved into one or more point sources by \emph{Chandra}. This decontamination is made possible by \emph{Chandra}'s on-axis sub-arcsecond PSF. This work is especially important given the upcoming launch of \emph{eROSITA} \citep{Merloni:2012a}. \emph{eROSITA}'s all-sky survey is expected to detect $\sim 10^5$ clusters out to redshifts $z>1$ \citep{Pillepich:2012a} and will have on-axis spatial resolution similar to that of \emph{XMM-Newton} and so will face the same challenges as \emph{XMM-Newton} in resolving point sources in distant clusters. 

The structure of the paper is as follows. In Section 2 we discuss the
sample selection and data preparation. Section 3 details the data
processing steps. Notes on individual clusters are given in Section 4. We discuss our results in Section 5. The conclusions are presented in Section 6. Throughout this paper we assume a WMAP9 cosmology of $H_0 = 70$ km/s/Mpc, $\Omega_{\Lambda} = 0.72$, and  $\Omega_{\m} = 0.28$ \citep{Hinshaw:2013a}. 


\begin{table*}
\begin{center}

\scalebox{0.85}{
\begin{tabular}{lcccccccccc}
  \hline
  XXLID & ObsID & Class & Class & $z$ & RA & Dec. &  F$_{60}$  & Chip  & Clean time\\
       &  & Willis & XXL & & (J2000) & (J2000) & ($10^{-14}\flux$)  & configuration & (ksec) \\
  \hline
XLSSC 072 & 18253 & C1 & C1 & 1.00 & 33.850 & -3.726 & 4.1$\pm$0.4 & ACIS-S & 9.9 \\
XLSSC 029 & 7185   & C1 & C1 & 1.05 & 36.017 & -4.225 &  3.2$\pm$0.3 & ACIS-S & 31.9 \\
XLSSC 005 & 18256 & C1 & C1 & 1.06 & 36.788 & -4.301 &  0.9$\pm$0.2 & ACIS-S & 10.9 \\
3XLSS J021825.9-045947 & 17306 &  & C1 &  1.13$^\diamond$ & 34.608 & -4.997 &  0.4$\pm$0.1 & ACIS-I$^\dagger$ & 50.8 \\
XLSSC 122 & 18263 & C1 & C1 & 1.99 & 34.433 & -3.759 &  1.3$\pm$0.3 & ACIS-S & 10.6 \\
\hline
XLSSC 048 & 18254 & C1 & C2 & 1.01 & 35.722 & -3.473 &  1.1$\pm$0.3 & ACIS-S & 9.4 \\
XLSSC 073 & 18255 & C1 & C2 & 1.03 & 33.744 & -3.506 &  0.7$\pm$0.3 & ACIS-S & 17.9 \\
3XLSS J022755.7-043119 & 20534 &  & C2 & 1.05$^\diamond$ & 36.982 & -4.522 &  0.3$\pm$0.3$^\ddagger$ & ACIS-S & 31.6 \\
3XLSS J021320.3-053411 & 20535 &  & C2 & 1.08$^\diamond$ & 33.334 & -5.570 &  0.1$^{+0.2}_{-0.1}$$^\ddagger$ & ACIS-S & 35.2 \\
XLSSC 203 & 17304 &  & C2 & 1.08 & 34.428 & -4.989 &  0.2$\pm$0.1 & ACIS-I$^\dagger$& 44.7 \\
XLSSC 634 & 11741 &  & C2 & 1.08 & 355.691 & -54.185 &  4.8$\pm$0.6 & ACIS-I$^\dagger$ & 62.7 \\
3XLSS J021325.0-042000 & 20536 &  & C2 & 1.20$^\diamond$ & 33.355 & -4.334 & 1.5$\pm$0.5$^\ddagger$ & ACIS-S & 9.9 \\
3XLSS J022005.5-050826 & 13374 & C2 & C2 & 1.65$^\diamond$ & 35.023 & -5.141 &  0.6$\pm$0.2$^\ddagger$ & ACIS-I & 75.7 \\
3XLSS J022418.4-043956 & 18262 & C2 & C2 & 1.67$^\diamond$ & 36.077 & -4.666 & 0.6$\pm$0.2$^\ddagger$ & ACIS-S & 11.9 \\
\hline
XLSSC 034 & 20538 &  & C3 & 1.04 & 35.372 & -4.099 & 2.1$\pm$0.9 & ACIS-S & 9.9 \\
3XLSS J022059.0-043922 & 18257 & C2 & C3 & 1.11$^\diamond$ & 35.246 & -4.657 &  0.9$\pm$0.3$^\ddagger$ & ACIS-S & 9.8 \\
XLSSC 046 & 18259 & C2 & C3 & 1.22 & 35.763 & -4.606 &  0.7$\pm$0.2 & ACIS-S & 20.8 \\
3XLSS J022351.3-041841 & 6390   &  & C3 & 1.27$^\diamond$ & 35.963 & -4.313 &  0.9$\pm$0.2$^\ddagger$ & ACIS-S$^\dagger$ & 10.8 \\
3XLSS J021700.4-034746 & 18260 & C2 & C3 & 1.54$^\diamond$ & 34.251 & -3.796 &  0.7$\pm$0.2$^\ddagger$ & ACIS-S & 9.9 \\
3XLSS J022812.3-043836 & 18261 & C2 & C3 & 1.67$^\diamond$ & 37.051 & -4.644 &  0.4$\pm$0.1$^\ddagger$ & ACIS-S & 9.6 \\
3XLSS J022554.3-045059 & 18264 & C2 & C3 & 2.24$^\diamond$ & 36.476 & -4.850 &  0.2$\pm$0.2$^\ddagger$ & ACIS-S & 21.7 \\
   \hline
\end{tabular}
}
\caption{\label{table:samplesummary}Summary of the cluster sample and \emph{Chandra} data. Column 1 is the cluster name; column 2 is the \emph{Chandra} ObsID; column 3 is the cluster class (see section 2) from \citet{Willis:2013a} or blank if the cluster is not part of that sample; column 4 is the cluster class from the updated XXL pipeline; column 5 is the redshift of the cluster (from XXL Paper \Rmnum{20} or for those not in that paper, the redshifts have not yet been published); columns 6 and 7 are the RA and Dec. coordinates of the cluster centre (from XXL Paper \Rmnum{20}); column 8 is the cluster flux in the 0.5 - 2 keV energy band measured in the 60$''$ cluster region using XXL data (those that are not included in XXL Paper \Rmnum{20} are marked with a $\ddagger$); column 9 is the CCD chip configuration for the observation  where a $\dagger$ means that the cluster fell off-axis in the observation - the off-axis distance is given in Section \ref{section.indivnotes}; column 10 is the cleaned \emph{Chandra} observation time. Redshifts that are photometric are marked with a $\diamond$
}
\end{center}
\end{table*}

\section{Sample and Data Preparation} \label{section.sample}

Our sample was initially constructed to comprise the 15 $z>1$ clusters and cluster candidates from the XMM-LSS survey (a $\sim$10 deg$^2$ precursor
to, and subset of XXL; \citealp{Willis:2013a}). The redshifts of two of those clusters (XLSS J022252.3-041647 and XLSSU J021712.1-041059) were subsequently revised to
be at z $<$ 1, so were dropped. Two of the remaining \citet{Willis:2013a} clusters had existing \emph{Chandra} archival data, the other 11 were targeted with new \emph{Chandra} snapshot observations. We subsequently expanded our sample to include a further four $z>1$ clusters detected in the wider XXL survey that have available Chandra data. The full 50 deg$^2$ XXL survey contains a further seven $z>1$ clusters for which we have been awarded Chandra observations, four of which have been observed and are included in this work, while the remaining three clusters have yet to be observed. Our final sample thus contains 21 $z>1$ clusters and candidates in total. 

The XXL source detection pipeline \textsc{Xamin} ranks clusters into classes 
(\citet{Pacaud:2006a}, \citet{Pacaud:2016a} - hereafter XXL Paper \Rmnum{2},  \citet{Faccioli:2018a} - also known as XXL Paper \Rmnum{24}). Galaxy cluster candidates are selected from the \textsc{Xamin} maximum likelihood outputs in \textsc{ext}, \textsc{ext\_stat} and \textsc{ext\_det\_stat}, which correspond to the extent, likelihood of extent, and detection
significance, respectively. A source is considered extended if it has measured \textsc{ext} greater than 5$''$ and \textsc{ext\_stat} greater than 15. The extended sources are then sorted into categories: the C1 class selects candidates with an \textsc{ext\_stat} greater than 33 and a \textsc{ext\_det\_stat} greater than 32; the C2 class comprises the remaining candidates. The C1 sample is expected to be mostly free of contamination by point sources. The C2 sample is expected to be about 50\% comprised of misclassified AGN, image artifacts and other spurious detections \citep{Pierre:2006a,Adami:2011a}, though it is worth noting that the contamination of the final C2 sample is likely to be significantly lower than this, as all cluster candidates are visually inspected, and obvious spurious sources are rejected. There exists a third class, the C3 sample, which consists of clusters known from optical/IR catalogues, that are associated with some X-ray emission that is too weak to be characterised \citep[see][ or XXL Paper \Rmnum{20}]{Pierre:2006a}. However, despite this, not all cluster candidates are expected to be genuine clusters: it is possible that in some cases where a cluster has been identified by XXL, there could just be a galaxy overdensity coincident with one or more AGN. The classifications were calibrated by simulations where the pipeline was run on previous \emph{XMM} observations with model clusters and randomly distributed AGN added \citep{Pacaud:2006a,Pacaud:2007a,Clerc:2012a}. These observations were restricted to low redshift clusters, and the purpose of this work is to extend this to lower signal-to-noise high redshift clusters which is more challenging due to the high redshift clusters often not being resolved, and there being bad supporting data.

The XXL analysis pipeline has been upgraded since the work reported in \citet{Willis:2013a}, leading to some changes in classification for individual objects (XXL Paper \Rmnum{24}). For the present analysis, we are using cluster classifications and properties consistent with those in the latest data release (XXL Paper \Rmnum{20}). Throughout this paper we often refer to the updated pipeline results, which are the results from \textsc{Xamin} consistent with the version used in XXL Paper \Rmnum{20}.

Our sample consists of five C1 clusters, nine C2 clusters and seven C3 clusters. Three C2 clusters (3XLSS J022755.7-043119, 3XLSS J021320.3-053411,3XLSS J021325.0-042000) and 1 C3 cluster are reported here for the first time. Table \ref{table:samplesummary} shows the properties of the clusters in our sample. The cluster flux in the 0.5 - 2 keV energy band measured in the 60$''$ cluster region using XXL data, F$_{60}$, reported in Table \ref{table:samplesummary} in column 8, was computed using a growth curve analysis as described in XXL Paper \Rmnum{2} (either taken from XXL Paper \Rmnum{20} or recomputed directly by us for objects not included in this paper). Two clusters (XLSSC 072 and XLSSC 029) are in the XXL 100 brightest galaxy cluster sample (XXL Paper \Rmnum{2}) and ten clusters (all C1s, 4 C2s - XLSSC 048, XLSSC 073, XLSSC 203, XLSSC 634 and 1 C3 - XLSSC 034) are in XXL Paper \Rmnum{20}.

The clusters in our \emph{Chandra} snapshot programme that were not covered by archival data were observed with the ACIS-S configuration with an exposure time designed to give a significant detection of a point source contributing $>$ 10\% of the 0.5 - 2 keV band XXL flux for C1s and spectroscopically confirmed C2s and $>$ 25\% for other cluster candidates. A minimum exposure time of 10 ks was imposed on all observations. The snapshot observations were not designed to detect significant emission from the ICM, although a borderline significant detection was expected in some cases. For those clusters already covered by archival data, two were in the ACIS-S configuration and four in the ACIS-I configuration (see Table \ref{table:samplesummary}). In some of the archived observations, the cluster fell relatively far from the optical axis, leading to a larger PSF
than for an on-axis observation, which sometimes caused complications in the analysis (see Section \ref{section.indivnotes}). 

All 21 clusters in our sample were analysed with the \textsc{ciao}\footnote{See \url{http://cxc.harvard.edu/ciao}} 4.9 software package and \textsc{caldb}\footnote{See \url{http://cxc.harvard.edu/caldb}} version 4.7.4 \citep{Fruscione:2006a}. The level 1 event files were reprocessed using the \texttt{chandra\_repro} tool following the standard data reduction threads\footnote{See \url{http://cxc.harvard.edu/ciao/threads/index.html}}. Periods of background flares were identified and removed using lightcurves analysed with the \texttt{deflare} tool. For observations taken in the ACIS-S configuration the cluster always fell on only the S3 chip, so  a lightcurve was extracted from only the S3 chip. For the observations in the ACIS-I configuration a lightcurve was extracted from the four front illuminated (FI) chips, CCD\_IDs I0-I3 (excluding any other chips in the observation). The CCDs not used for the lightcurve filtering were discarded from the rest of the analysis. 

\begin{figure*}
\begin{center}
{\includegraphics[width = 51mm]{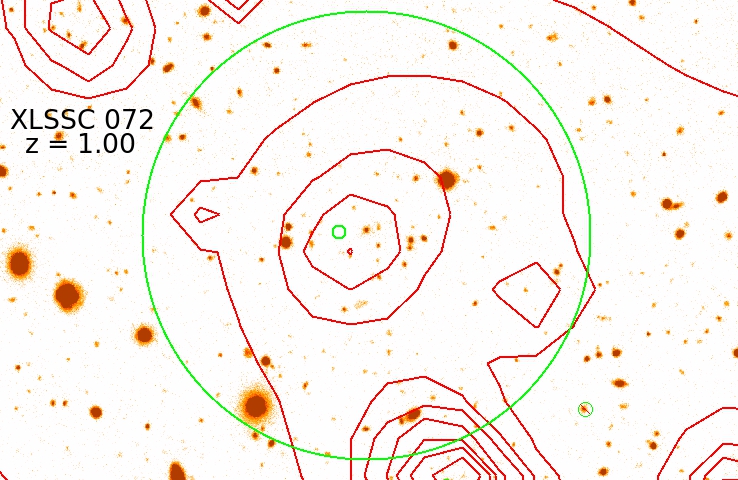}}
{\includegraphics[width = 51mm]{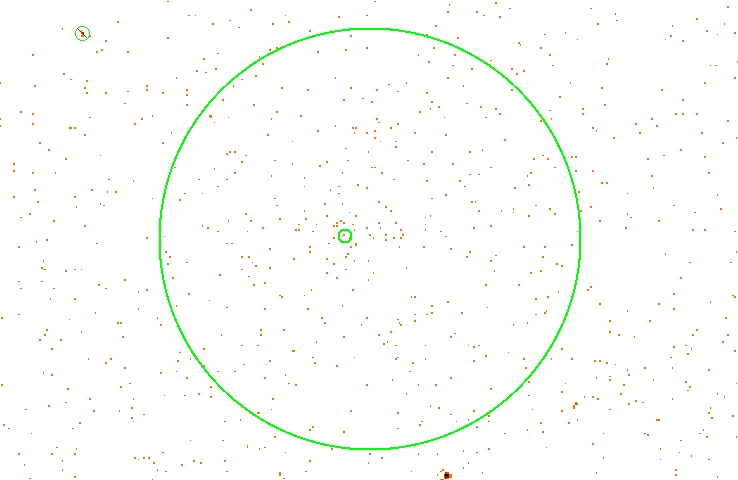}}
{\includegraphics[width = 51mm]{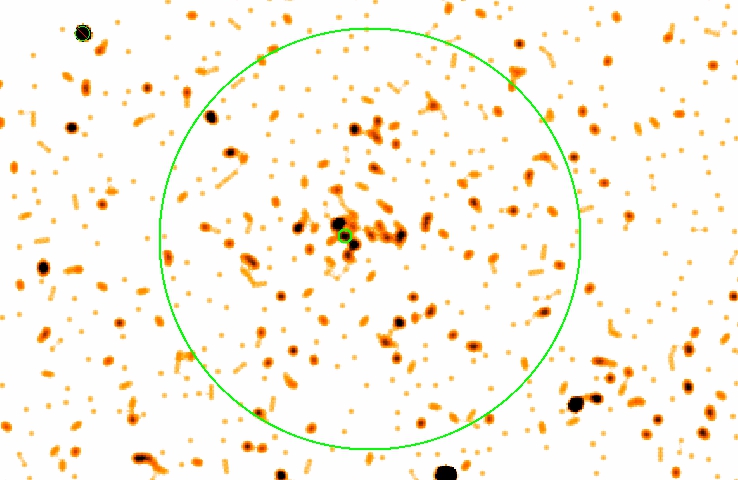}}
{\includegraphics[width = 51mm]{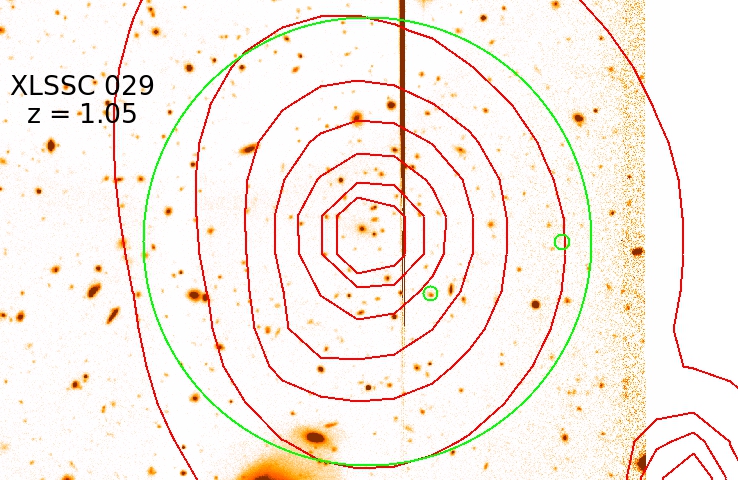}}
{\includegraphics[width = 51mm]{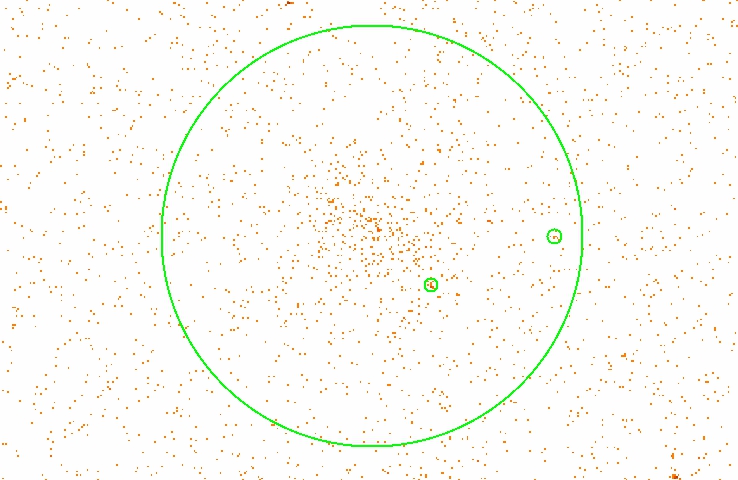}}
{\includegraphics[width = 51mm]{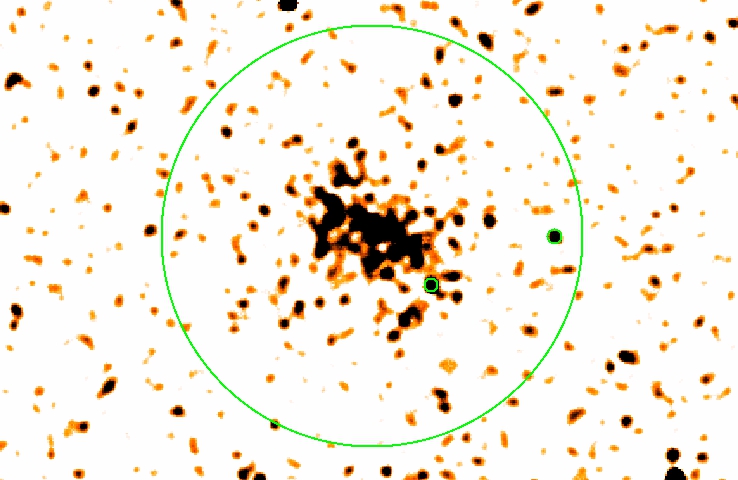}}
{\includegraphics[width = 51mm]{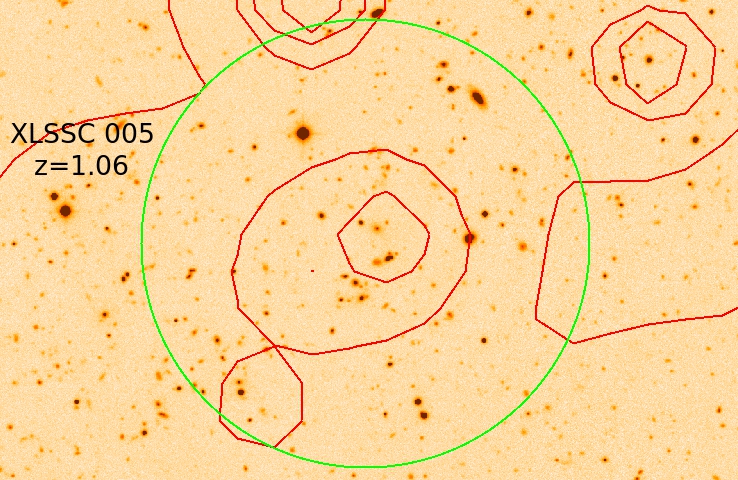}}
{\includegraphics[width = 51mm]{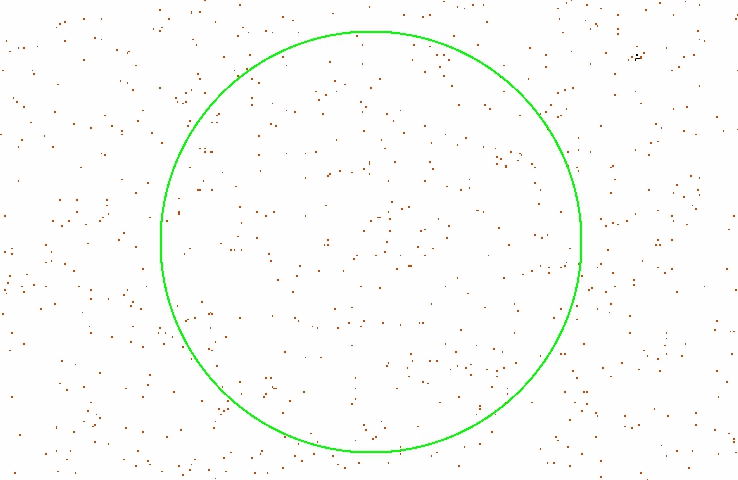}}
{\includegraphics[width = 51mm]{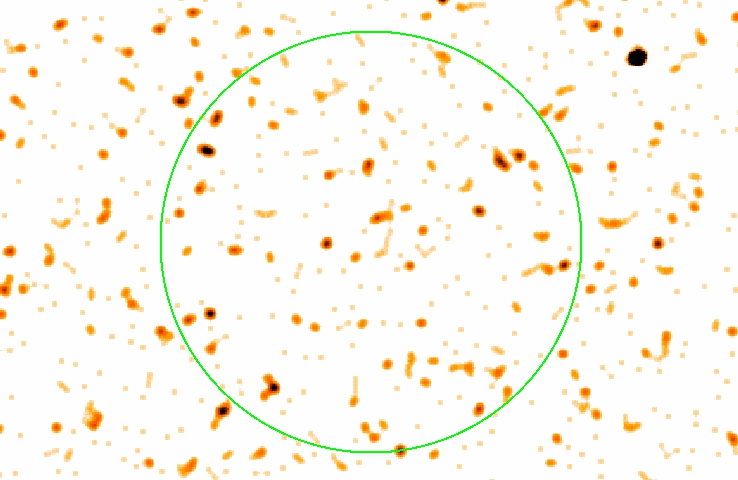}}
{\includegraphics[width = 51mm]{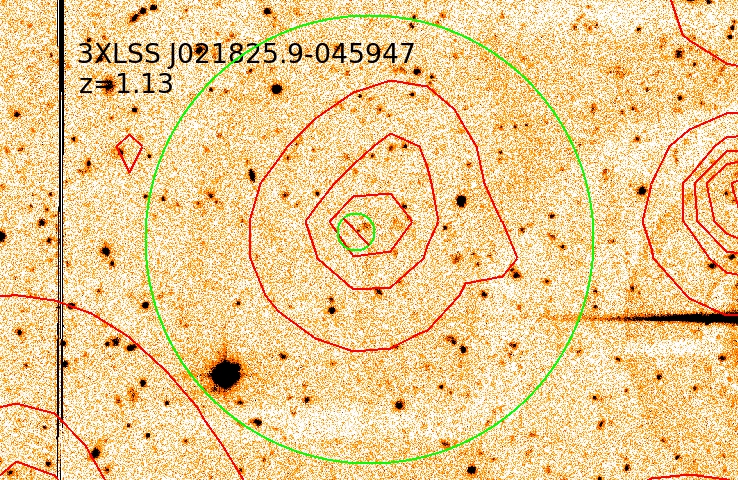}}
{\includegraphics[width = 51mm]{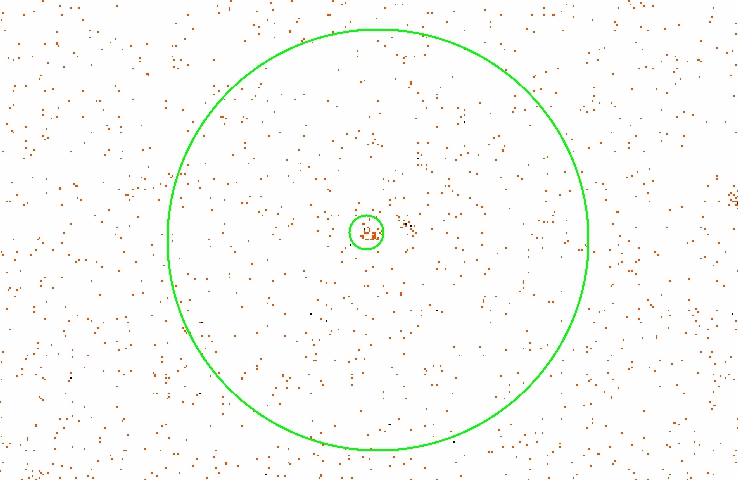}}
{\includegraphics[width = 51mm]{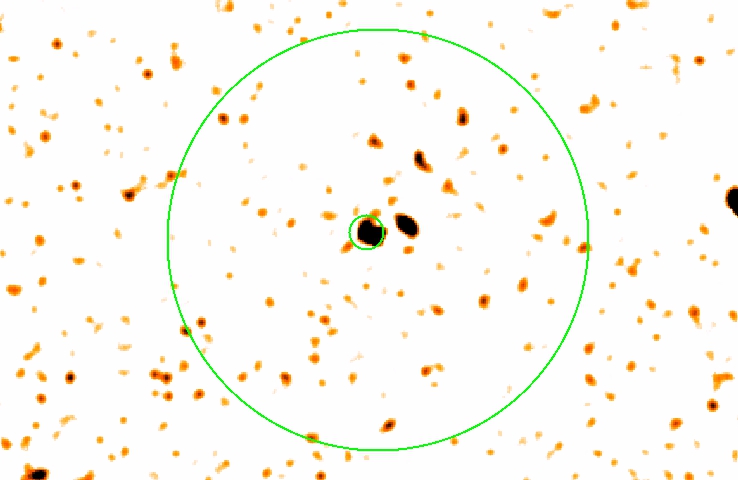}}
{\includegraphics[width = 51mm]{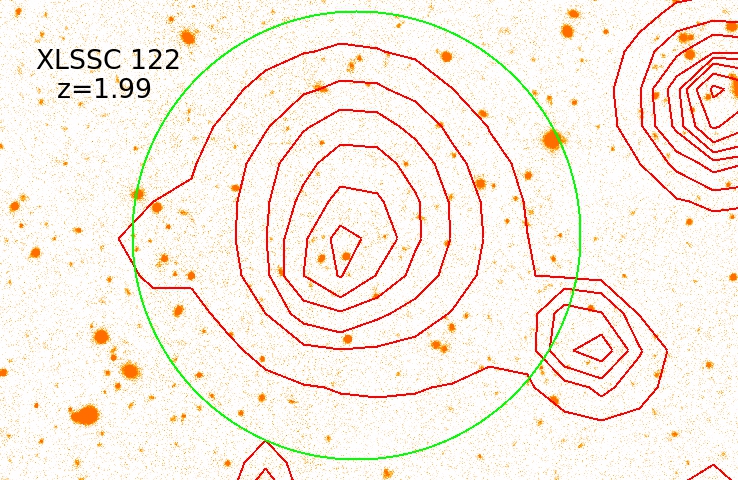}}
{\includegraphics[width = 51mm]{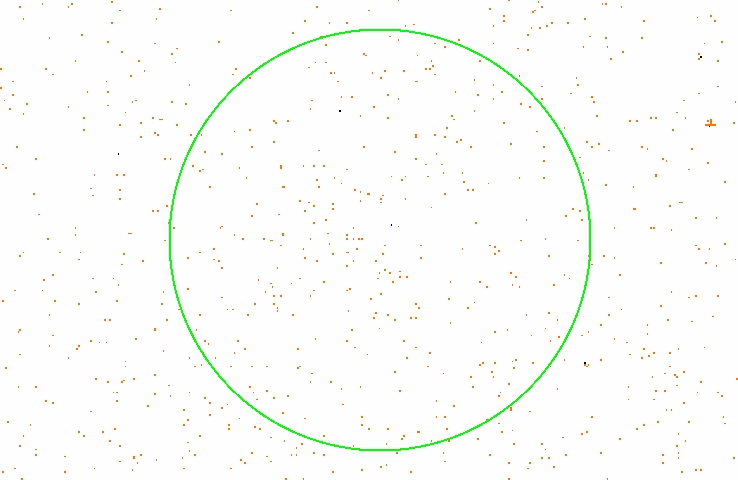}}
{\includegraphics[width = 51mm]{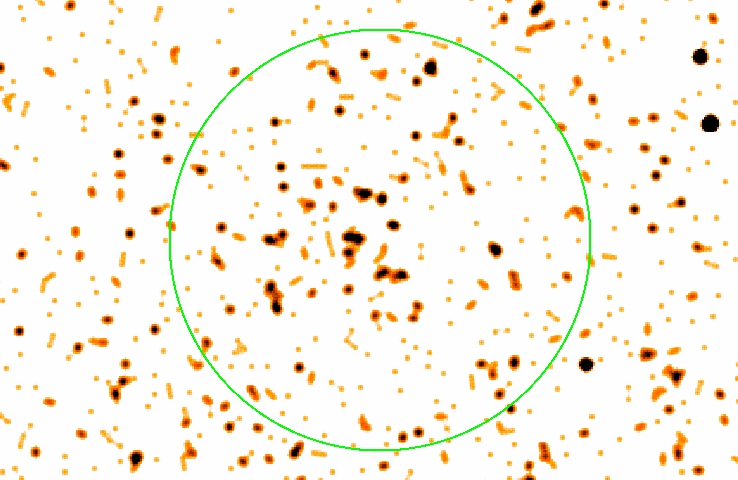}}
\caption[]{A comparison of the optical image with the \emph{XMM-Newton} contours from the 0.5 - 2 keV band (red) superimposed (left) and the raw and smoothed (using a Gaussian with $\sigma \sim 2.5''$) \emph{Chandra} (centre and right, respectively) images for all C1 clusters. All optical images are i-band images from the CFHTLS except for 3XLSS J021825.9-045947 which is r-band. \emph{Chandra} images are in the 0.3 - 8.0 keV band. The green circle is the same in all images and is of radius 60$''$ and centred on the cluster centre. Point sources within 60$''$ of the cluster centre are marked by the smaller green circles in all images. In the raw \emph{Chandra} images, if a \emph{Chandra} point source was detected in XXL then it is circled in red.}
	\label{fig:threeimagesC1}
\end{center}
\end{figure*}

In Figures \ref{fig:threeimagesC1}, \ref{fig:threeimagesC2} and \ref{fig:threeimagesC3} we show optical and \emph{Chandra} images for the C1, C2 and C3 clusters respectively.


\begin{figure*}
\begin{center}
{\includegraphics[width = 51mm]{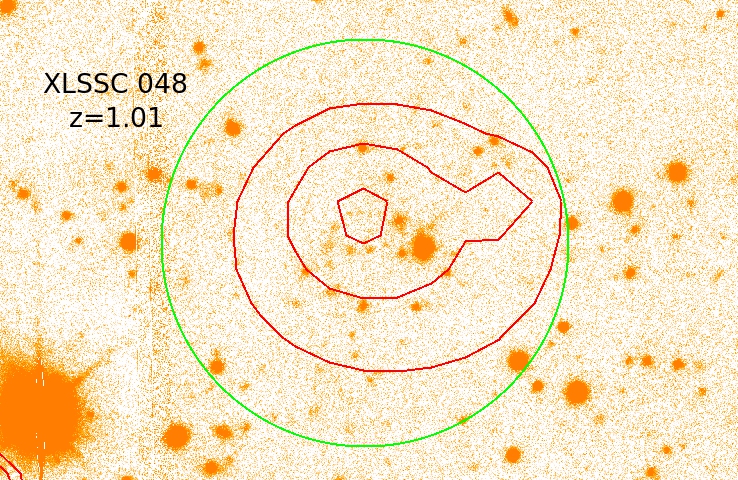}}
{\includegraphics[width = 51mm]{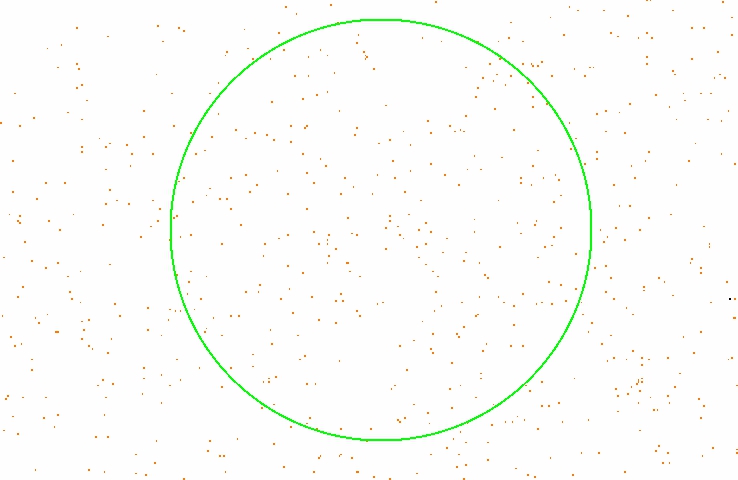}}
{\includegraphics[width = 51mm]{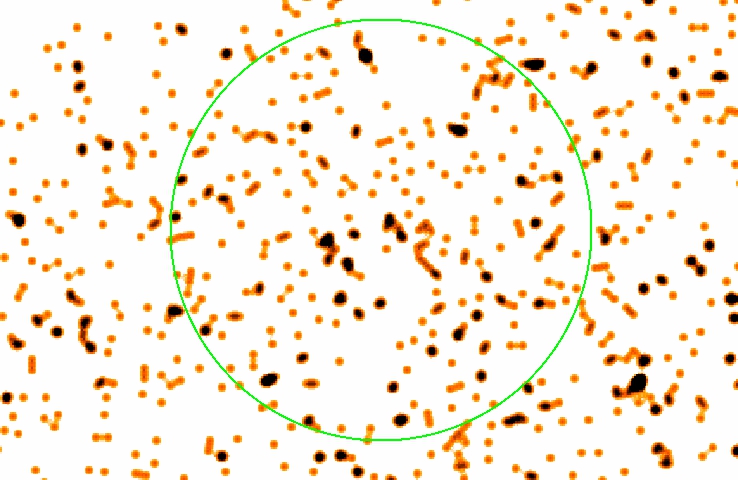}}
{\includegraphics[width = 51mm]{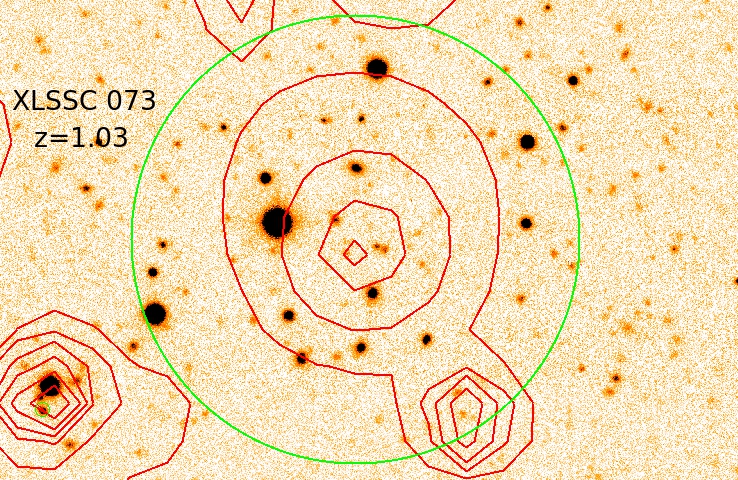}}
{\includegraphics[width = 51mm]{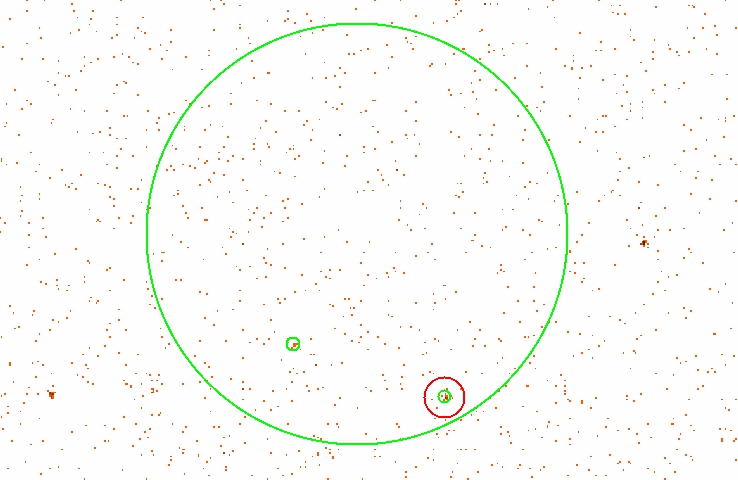}}
{\includegraphics[width = 51mm]{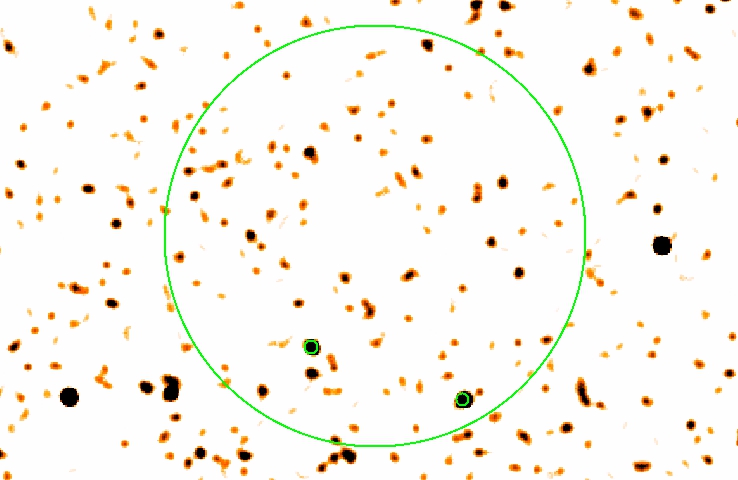}}

{\includegraphics[width = 51mm]{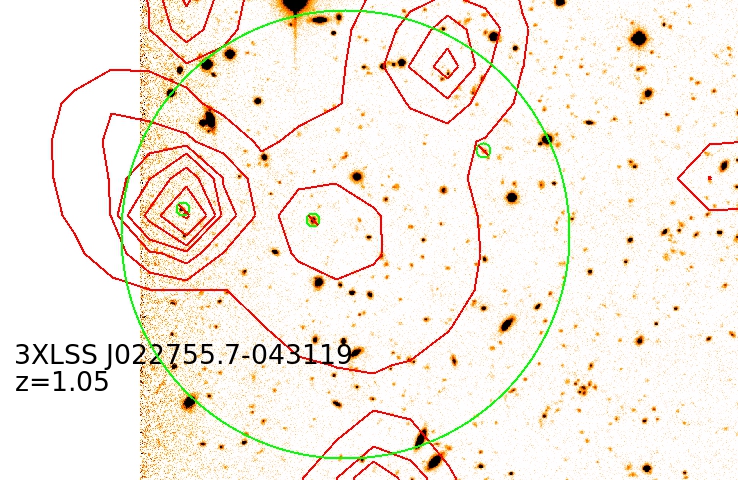}}
{\includegraphics[width = 51mm]{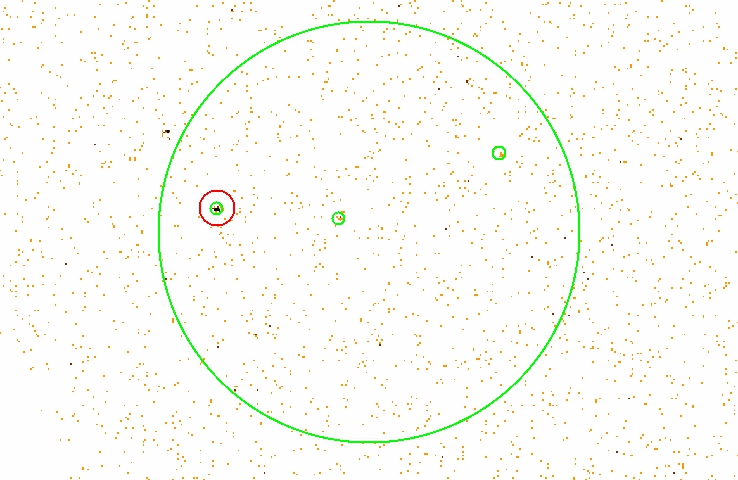}}
{\includegraphics[width = 51mm]{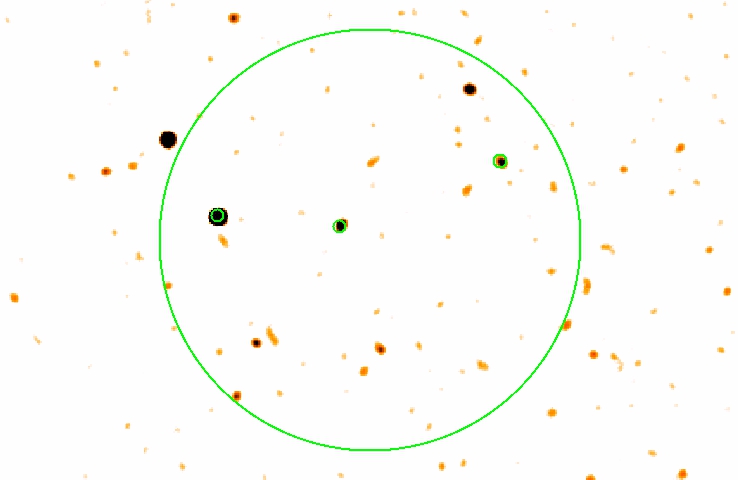}}
{\includegraphics[width = 51mm]{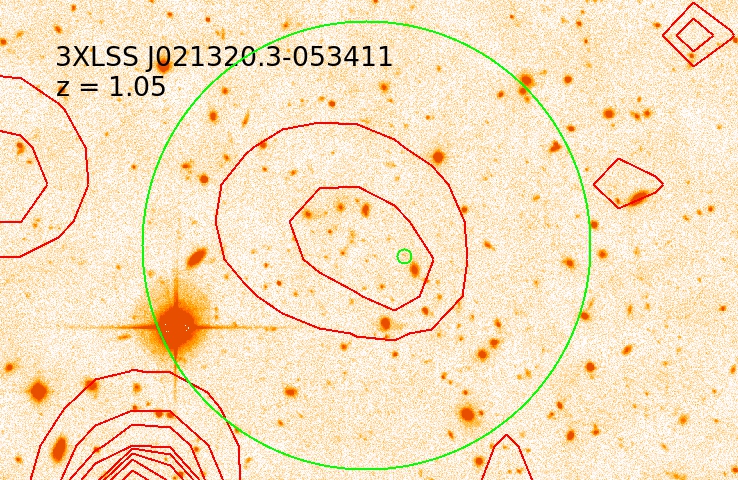}}
{\includegraphics[width = 51mm]{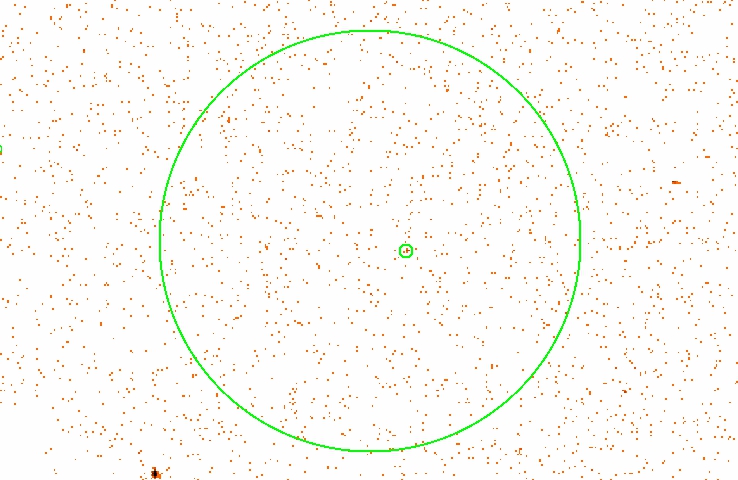}}
{\includegraphics[width = 51mm]{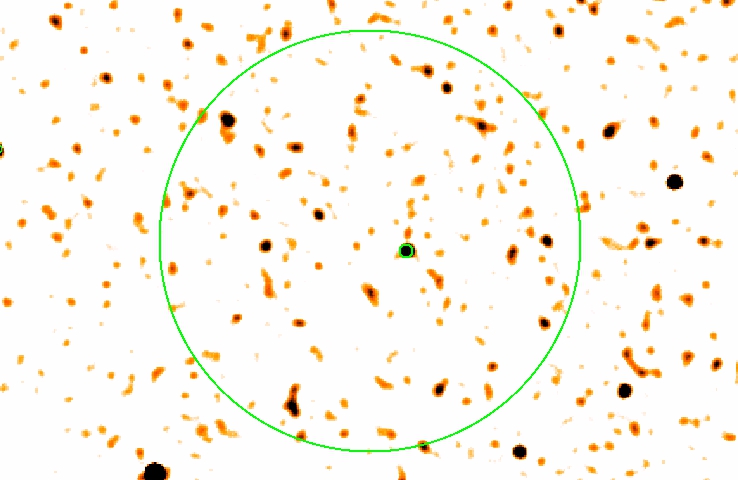}}

{\includegraphics[width = 51mm]{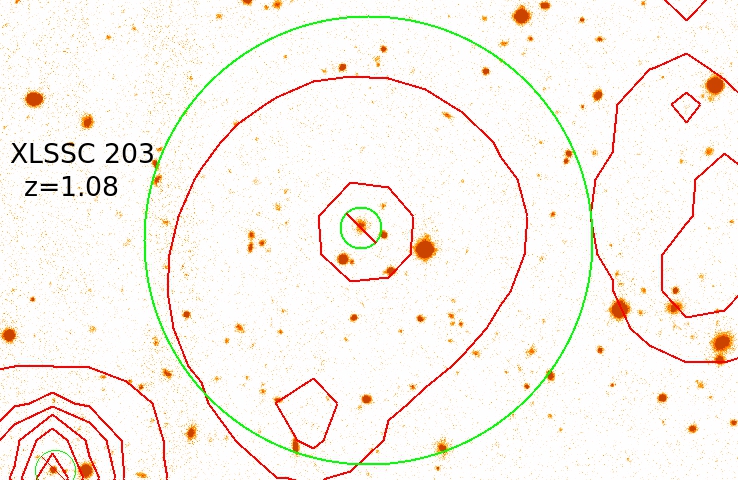}}
{\includegraphics[width = 51mm]{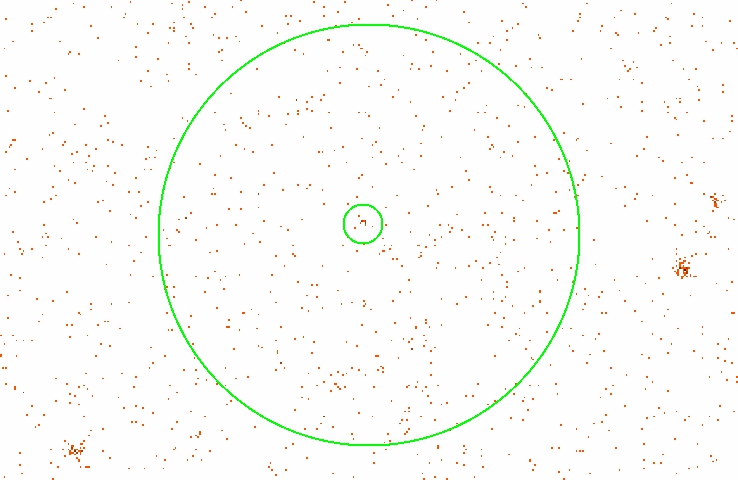}}
{\includegraphics[width = 51mm]{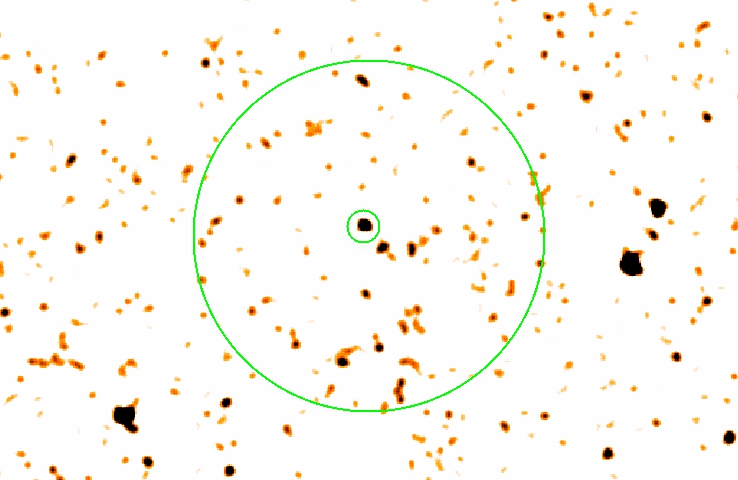}}
{\includegraphics[width = 51mm]{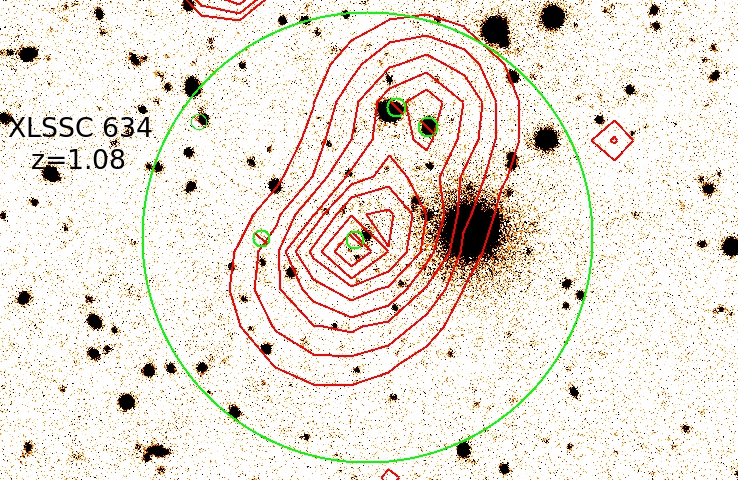}}
{\includegraphics[width = 51mm]{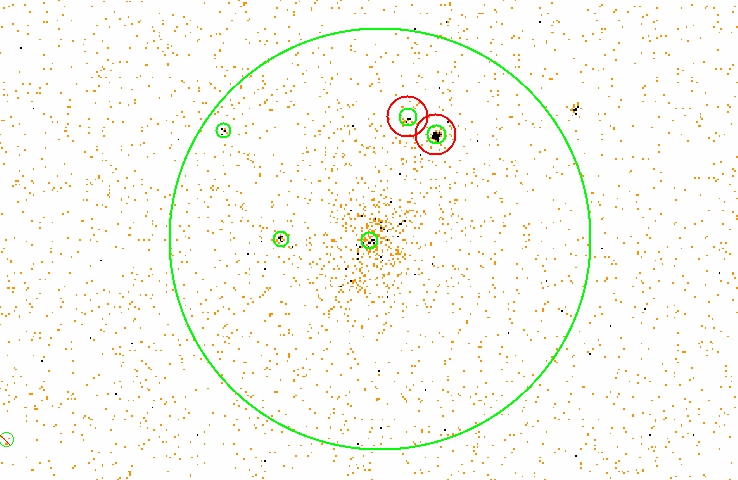}}
{\includegraphics[width = 51mm]{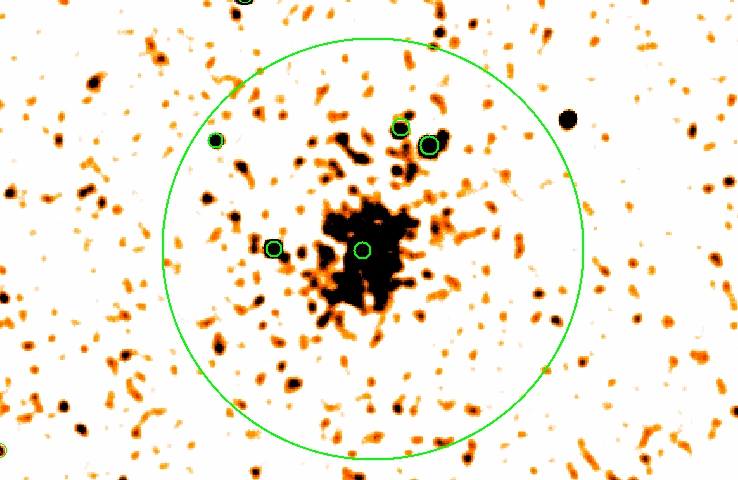}}

{\includegraphics[width = 51mm]{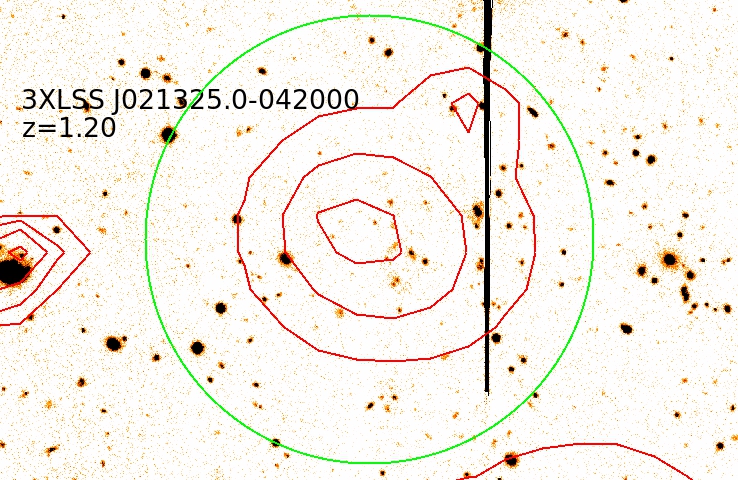}}
{\includegraphics[width = 51mm]{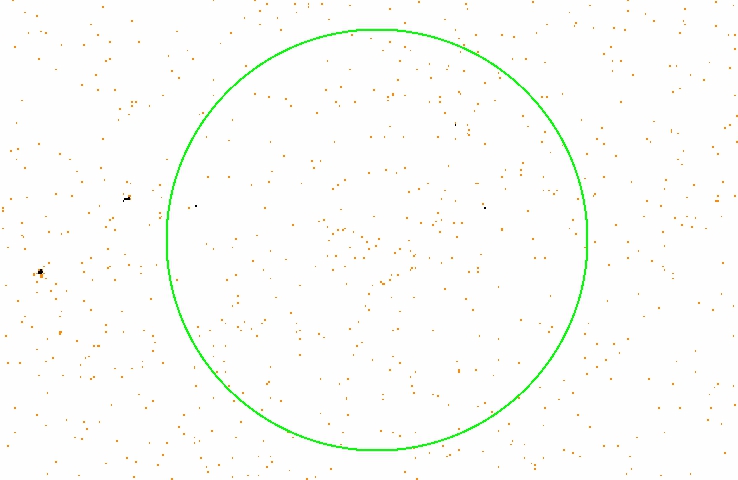}}
{\includegraphics[width = 51mm]{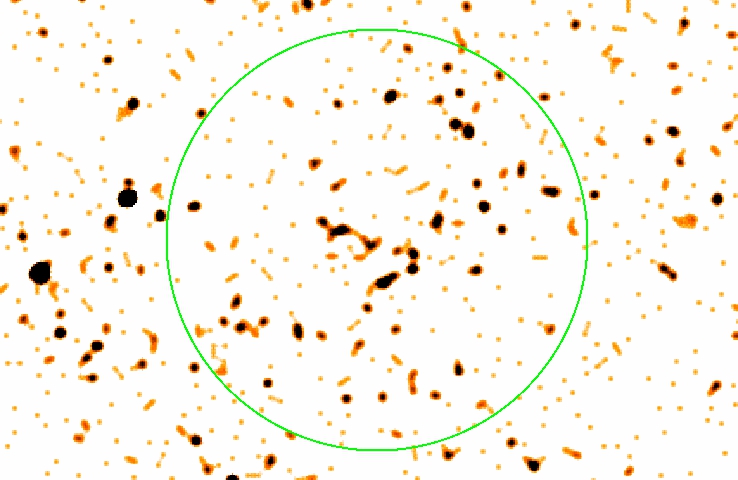}}
\caption[]{Same as Figure \ref{fig:threeimagesC1} but for all C2 clusters. All optical images are i-band images from the CFHTLS except for XLSSC 203 which is r-band and XLSSC 073 which is g-band. }
	\label{fig:threeimagesC2}
\end{center}
\end{figure*}

\renewcommand{\thefigure}{\arabic{figure}}
\addtocounter{figure}{-1}

\begin{figure*}
\begin{center}
{\includegraphics[width = 51mm]{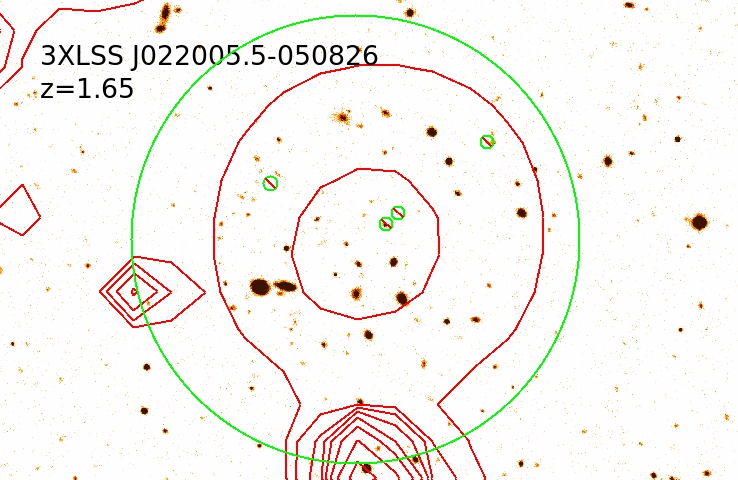}}
{\includegraphics[width = 51mm]{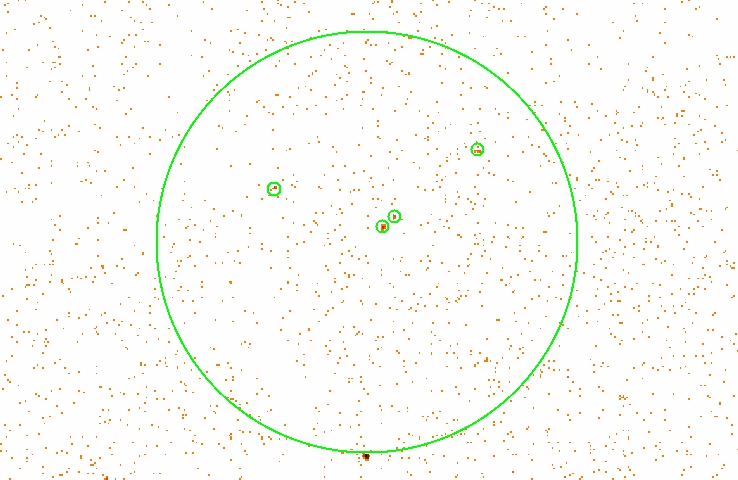}}
{\includegraphics[width = 51mm]{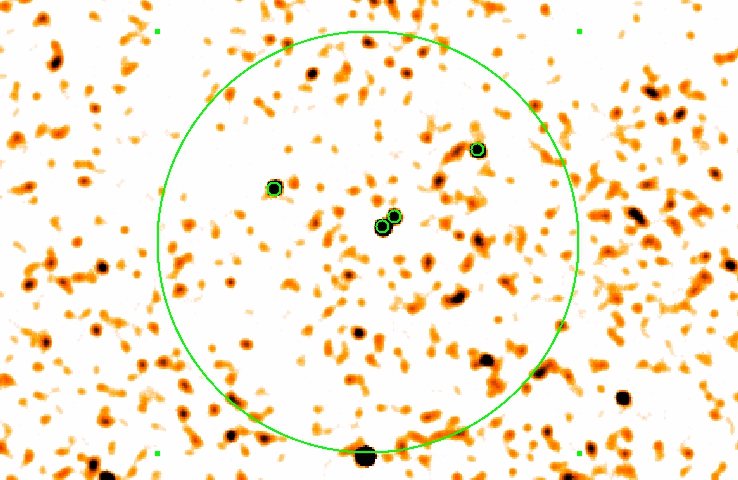}}
{\includegraphics[width = 51mm]{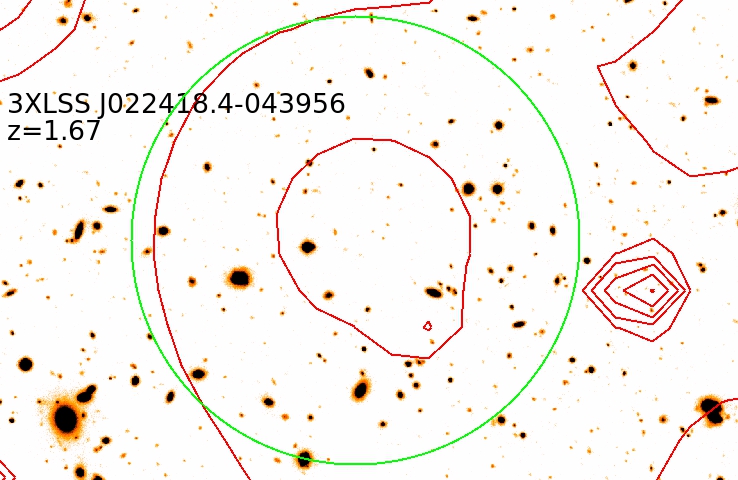}}
{\includegraphics[width = 51mm]{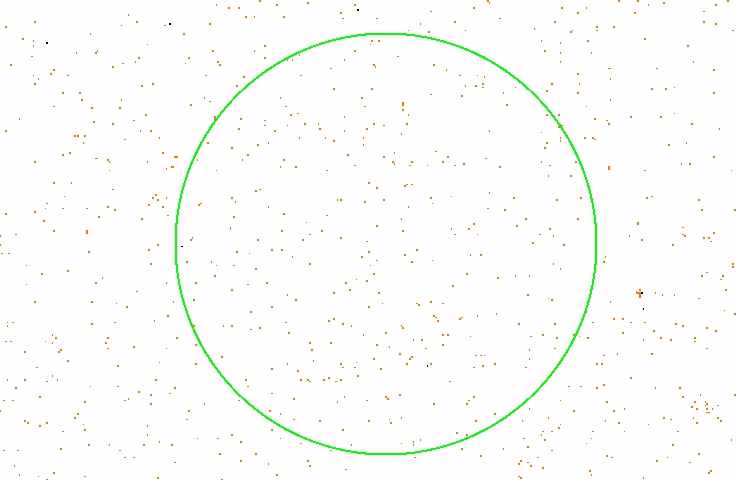}}
{\includegraphics[width = 51mm]{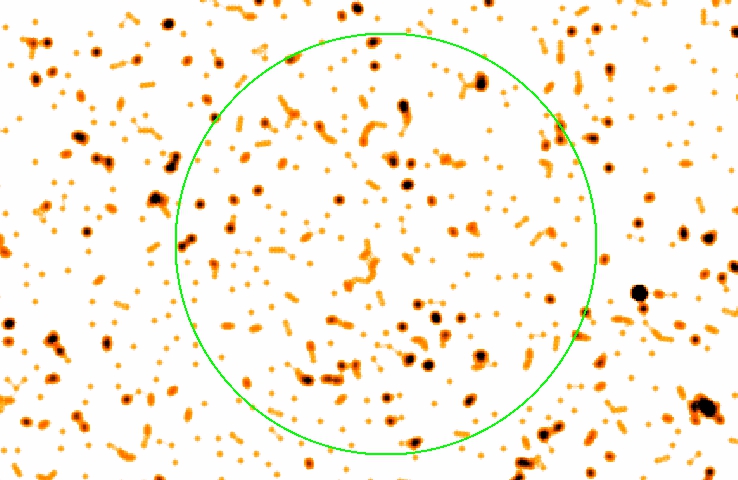}}
\caption{\emph{- continued}}
\end{center}
\end{figure*}

\renewcommand{\thefigure}{\arabic{figure}}

\begin{figure*}
\begin{center}
\includegraphics[width = 51mm]{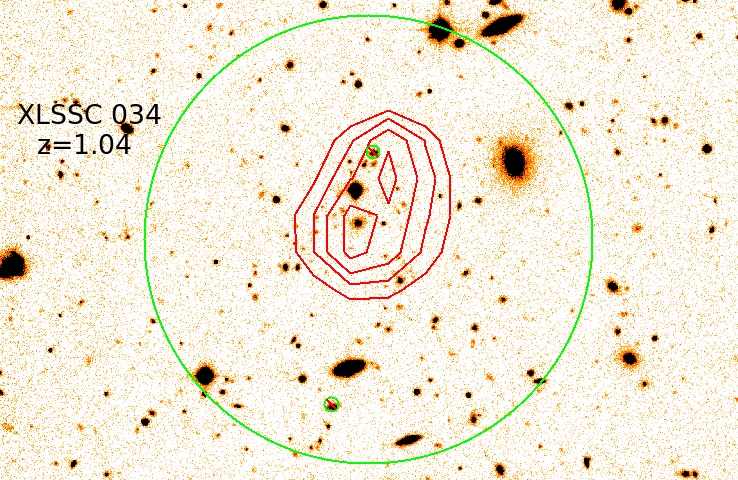}
\includegraphics[width = 51mm]{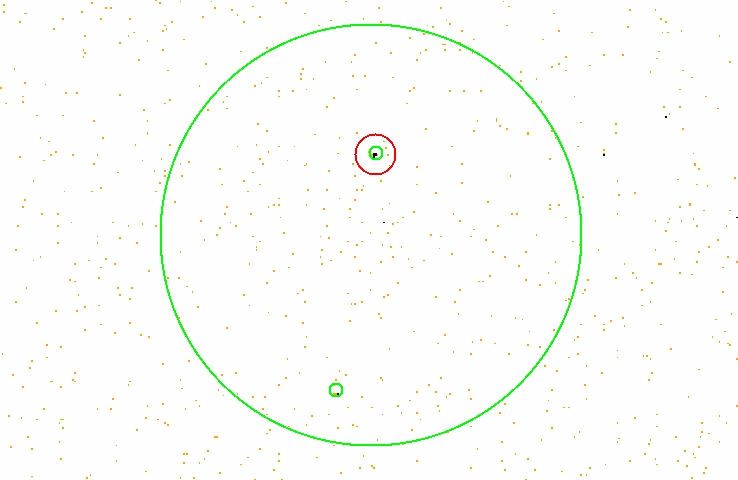}
\includegraphics[width = 51mm]{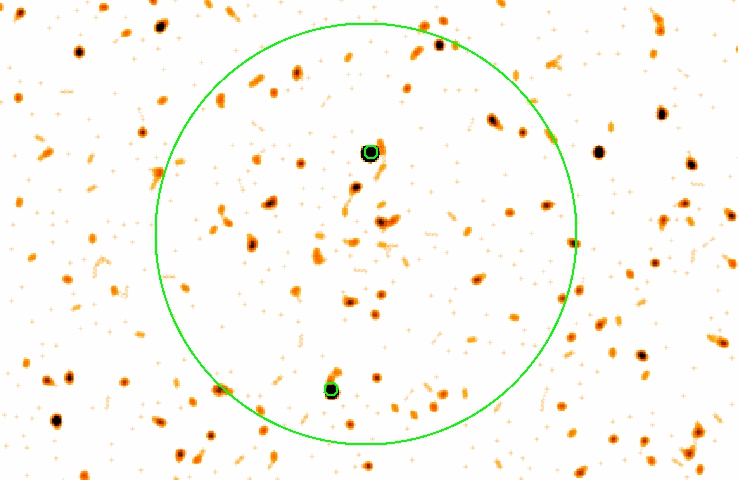}
\includegraphics[width = 51mm]{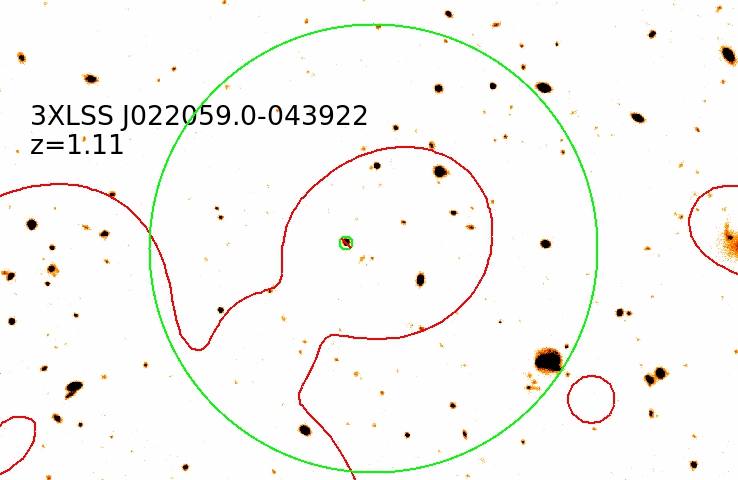}
\includegraphics[width = 51mm]{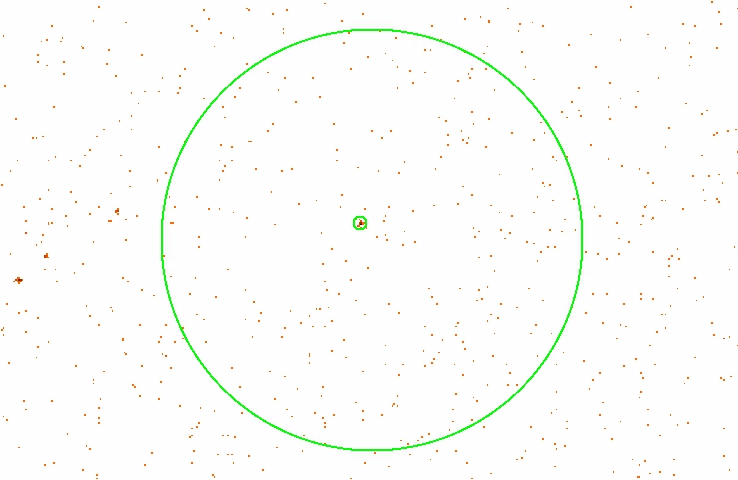}
\includegraphics[width = 51mm]{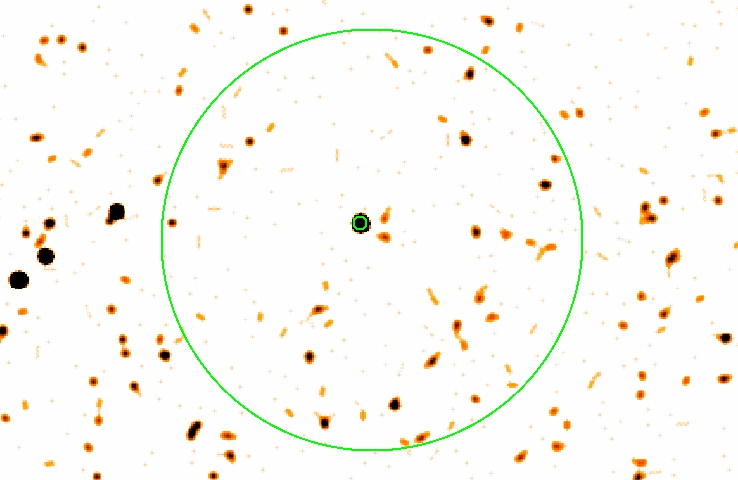}
\includegraphics[width = 51mm]{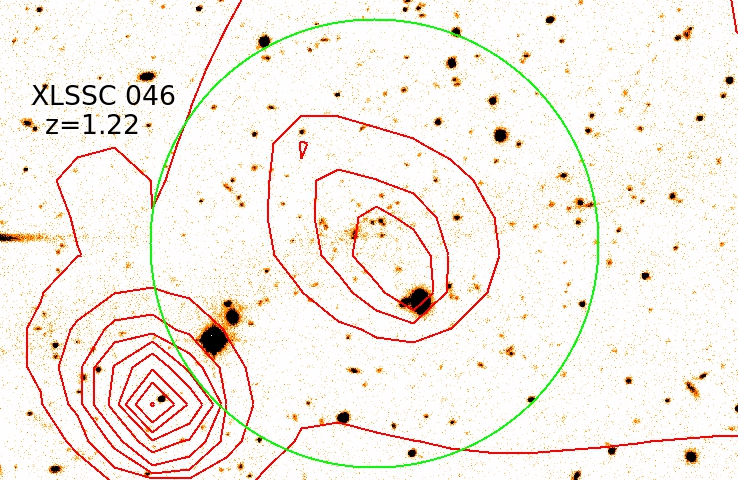}
\includegraphics[width = 51mm]{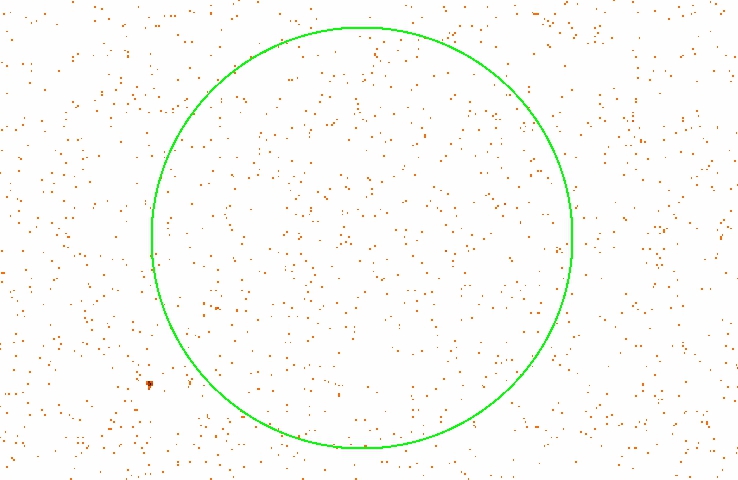}
\includegraphics[width = 51mm]{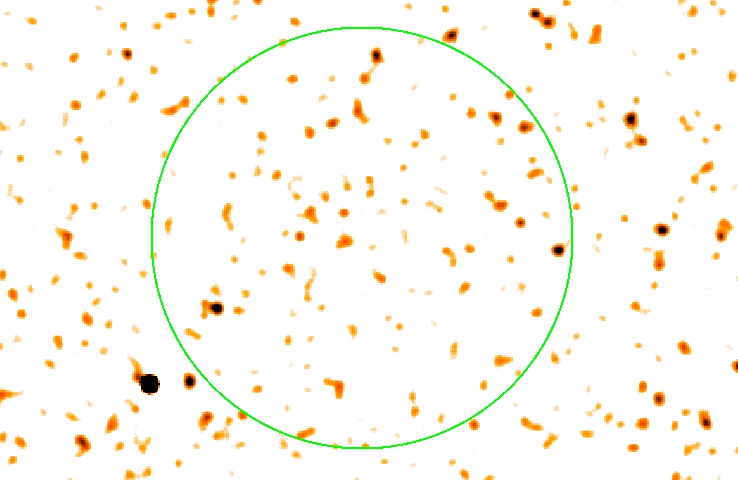}
\includegraphics[width = 51mm]{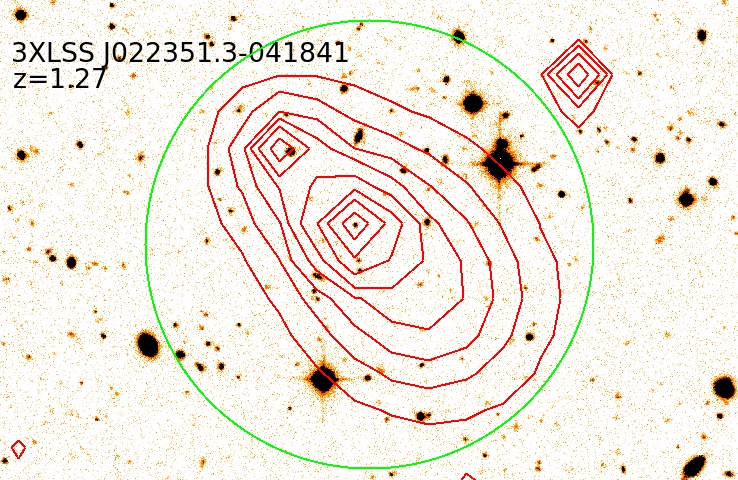}
\includegraphics[width = 51mm]{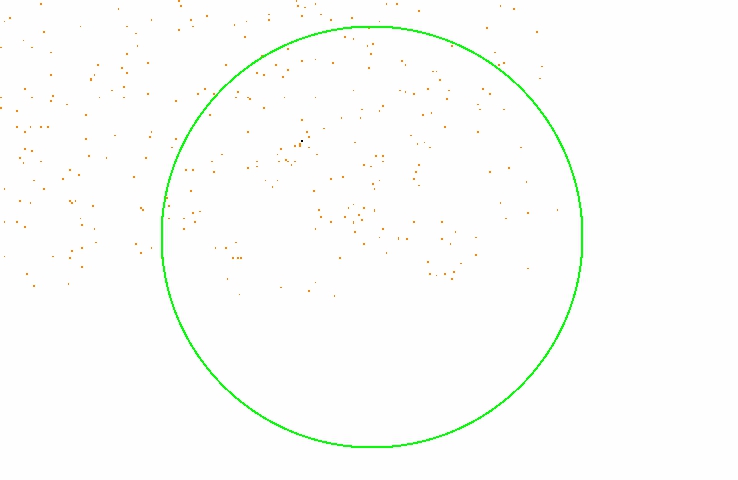}
\includegraphics[width = 51mm]{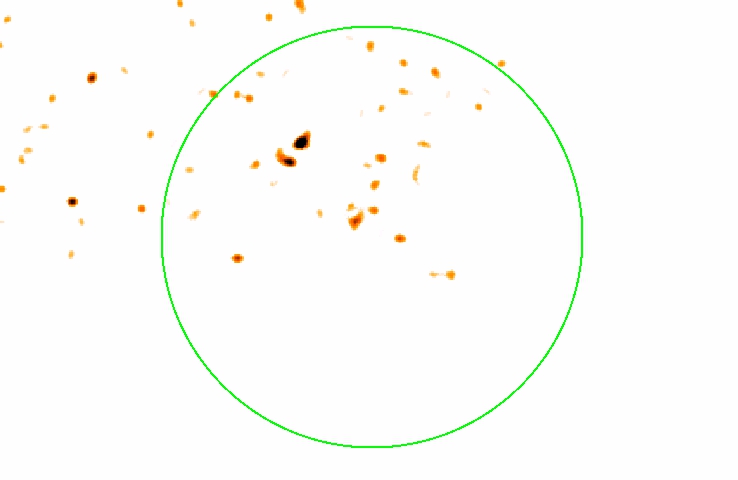}
\includegraphics[width = 51mm]{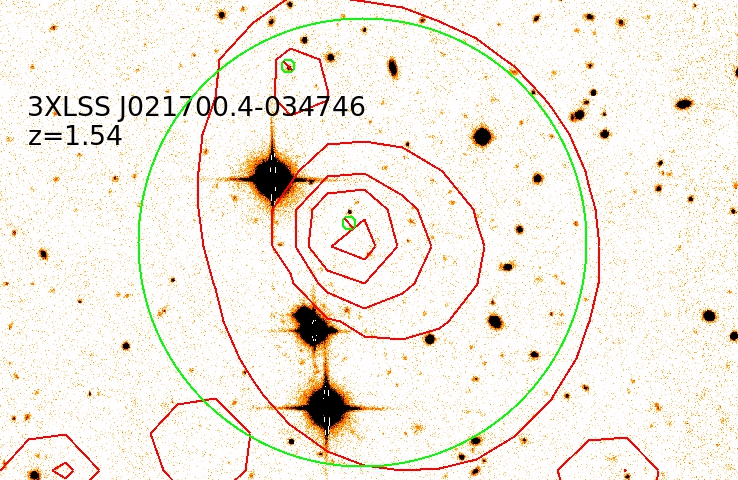}
\includegraphics[width = 51mm]{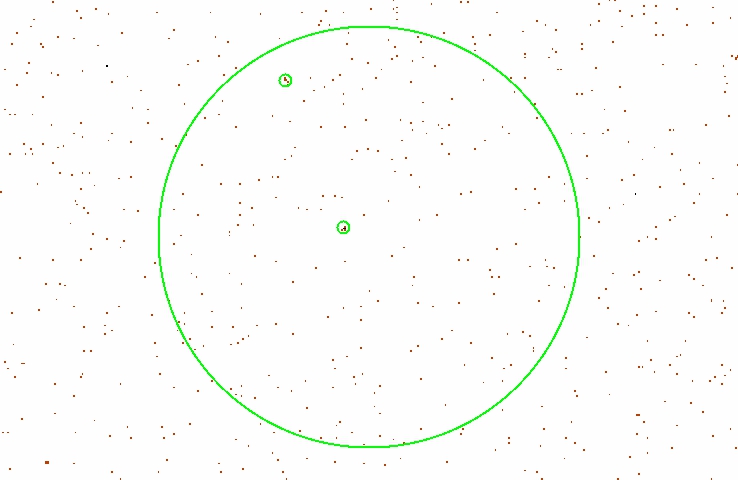}
\includegraphics[width = 51mm]{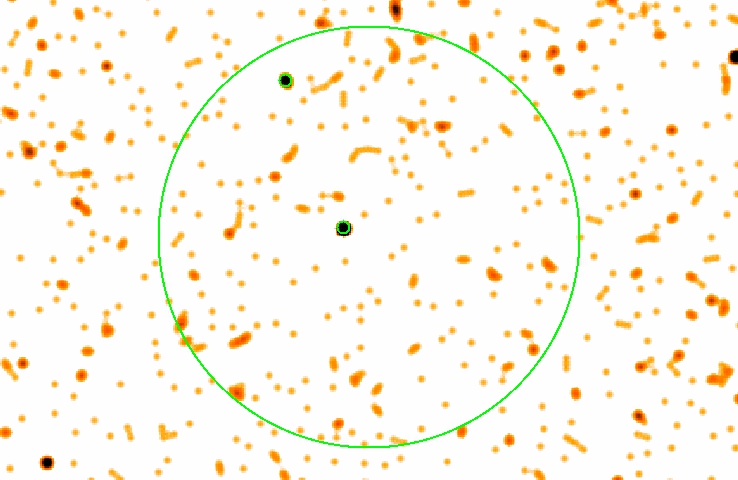}
\includegraphics[width = 51mm]{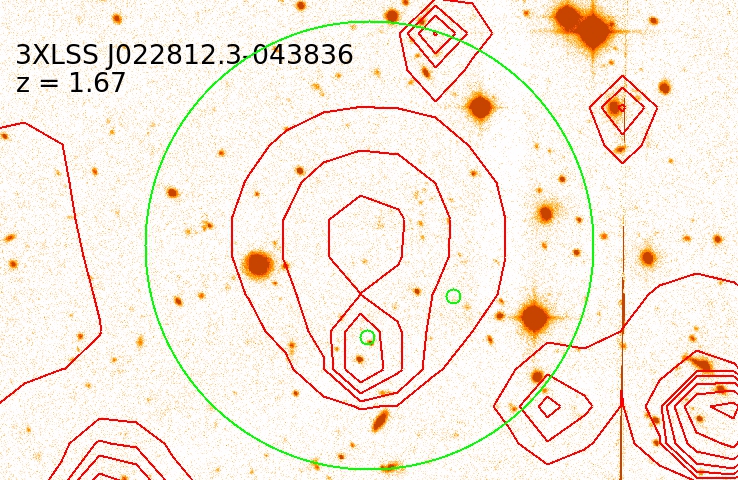}
\includegraphics[width = 51mm]{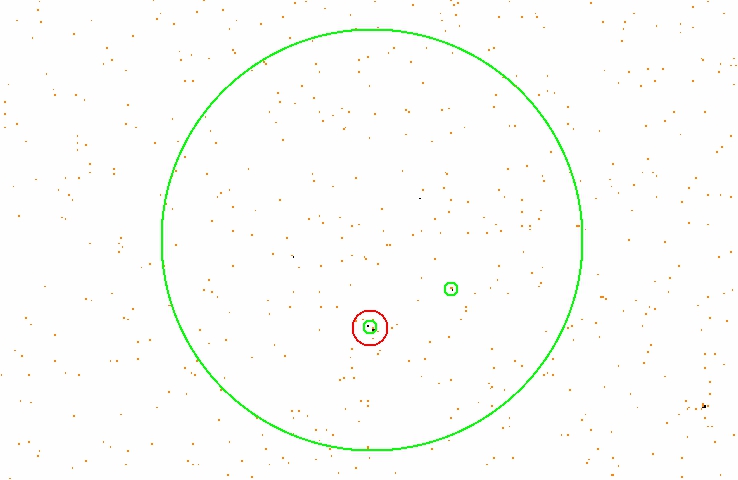}
\includegraphics[width = 51mm]{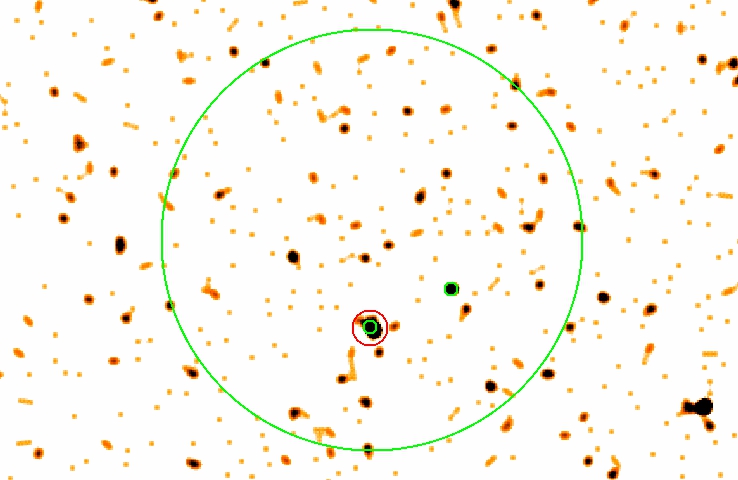}
\includegraphics[width = 51mm]{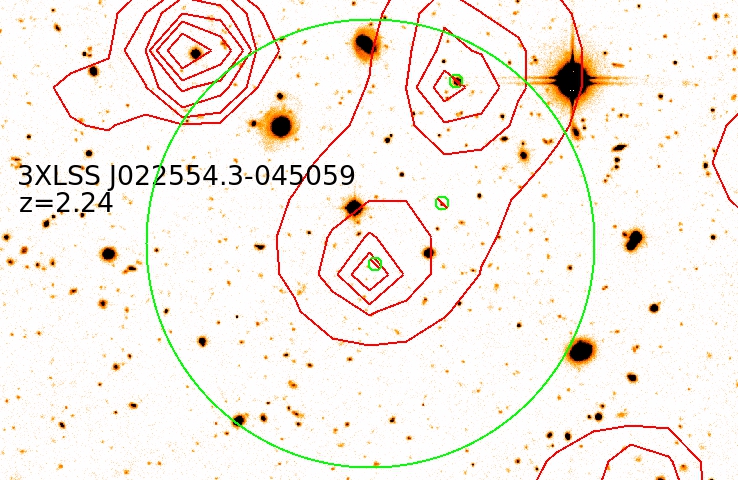}
\includegraphics[width = 51mm]{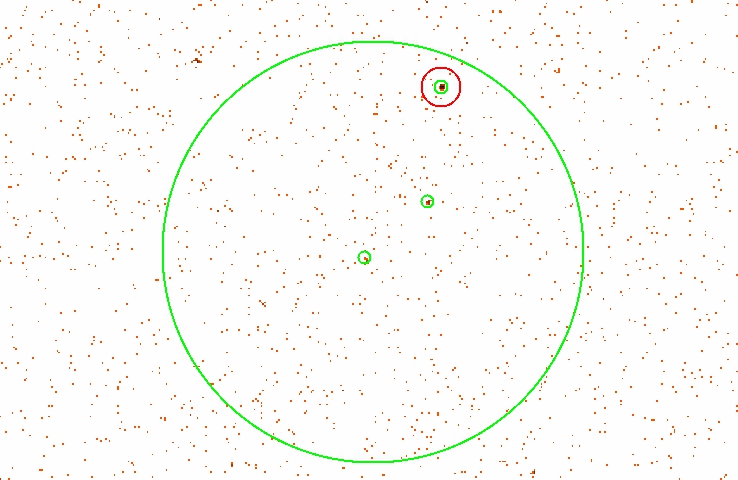}
\includegraphics[width = 51mm]{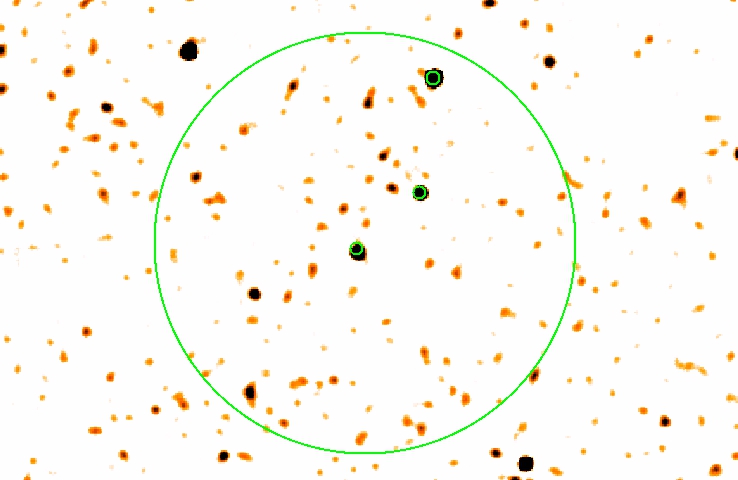}
\caption[]{Same as Figure \ref{fig:threeimagesC1} but for all C3 clusters. All optical images are i-band images from the CFHTLS.}
	\label{fig:threeimagesC3}
\end{center}
\end{figure*}

%
%


\section{Data Processing} \label{section.dataprocessing}

The main focus of our analysis is both to obtain flux constraints for detected sources, and to determine upper limits for possible sources that were not detected. For source detection we use the \textsc{ciao} \texttt{wavdetect} tool, and for photometry the \textsc{ciao} \texttt{srcflux} tool was used. The \texttt{srcflux} tool uses a Bayesian method to compute the background-marginalised posterior probability distribution of the source flux. \texttt{srcflux} has three possible outcomes: a ``good measurement'' where the probability distribution function (PDF) is not truncated at zero for the confidence interval specified, so the lower limit is given as well as the most probable flux and upper limit; ``pdf truncated at zero'' where the most probable flux and upper limit are given, but the lower limit is not given as the PDF is truncated at zero for the confidence interval specified; ``mode of zero'' where the most probable flux is zero and a lower limit is therefore not given, but an upper limit is still given. 

In the following section we describe the detection and photometry of point sources in the \emph{Chandra} data in or projected onto the cluster regions. We assume that all point sources detected are AGN, as AGN vastly outnumber any other contaminating point sources at this depth - the possibility that they could be X-ray bright stars is $\sim$3\% (\citet{Galametz:2009a},\citet{Chiappetti:2018a} - also known as XXL Paper \Rmnum{27}). For several clusters, point sources were detected in these regions by the XXL pipeline and excluded from the XXL cluster flux measurements. Since the goal of our analysis is to estimate the effects of AGN that were unresolved by \emph{XMM} we do not include the point sources that were detected by XXL in the main body of this paper. These sources are detailed in Table \ref{table:indivptsrcfluxessummary}.

\subsection{Point Source Detection and Flux Calculation} \label{subsection.ptsrcdetnandflux}

For the purpose of point source detection, images and the appropriate exposure maps were produced in the 0.3 - 8 keV band \citep{Kim:2007a}. The \textsc{ciao} \texttt{wavdetect} tool was used to search for point sources in these images. The \texttt{scales} parameter was set as $(\sqrt{2})^n$ with n $=$ 0 - 8 and the \texttt{sigthresh} parameter was set to 1 $\times$ $10^{-6}$ such that there will be $\sim$4 false-positive source detections per image for the 4 FI chips in the ACIS-I observations and $\sim$1 for the S3 chip in the ACIS-S observations. Since we are considering only the 60$''$ region around the cluster, the false positive rate will be $\sim$0.05 false-positive source detections per cluster, corresponding to $\sim$1 false positive in the full sample of
clusters. 
The detection limit corresponds to $\sim$5 photons from the source aperture in \texttt{wavdetect}.

In some cases where the cluster fell off-axis, due to the observation being from pre-existing \emph{Chandra} data not specifically designed to observe the cluster, there was ambiguity as to whether a detected source was a point source or ICM emission. There were also cases where no source was detected by \texttt{wavdetect} but a visual inspection suggested a possible point source in or projected onto the cluster region. In order to be conservative in our classification of whether point sources were present, we flagged as possible point sources any regions within 60$''$ of the cluster centre that possessed either (i) at least 4 counts in a single pixel, or (ii) at least 6 counts in a 1$''$ circle with at least one pixel containing 2 or more counts. This formalised our visual inspection enabling us to apply it to simulated images when determining upper limits as described below.

Multi-wavelength data were used to assist the classification of these possible point sources, and details for each are given in Section \ref{section.indivnotes}. For the optical band we used the Canada-France-Hawaii Telescope Legacy Survey (CFHTLS) data  for XXL-N\footnote{\url{http://www.cfht.hawaii.edu/Science/CFHTLS}}. These images were taken with the wide field optical imaging camera MegaCam, a 340 Megapixel camera with a 1$'$ by 1$'$ field of view. For XLSSC 634 in XXL-S the image was taken from the BCS survey \citep{Desai:2012a} with the Mosaic2 imager on the Blanco 4m telescope\footnote{\url{http://www.ctio.noao.edu/noao/node/9}}. For radio data we used the NRAO VLA Sky Survey \citep{Condon:1998a} and \citet{Tasse:2008a} for the XXL-N field and used Australia Telescope Compact Array (ATCA) data 
(\citet{Smolcic:2016a} - also known as XXL Paper \Rmnum{11}, \citet{Butler:2017a} - also known as XXL Paper \Rmnum{18}) for the XXL-S field (for XLSSC 634). We define an optical or radio source as a likely counterpart to a \emph{Chandra} detected point source if it falls within 2$''$ of the \emph{Chandra} detected point source coordinates. 

Fluxes were then measured for all point sources detected within 60$''$ of the cluster centre (as in XXL Paper \Rmnum{20}), assuming a power law model with $\Gamma=$ 1.7, consistent with the modelling used in other XXL papers 
(\citet{Fotopoulou:2016a} - also known as XXL Paper \Rmnum{6}, XXL Paper \Rmnum{27}); however, since we are measuring the flux in a relatively narrow band (compared to the full \emph{Chandra} bandpass), without needing to extrapolate, and with too few counts to fit the spectral index, the exact choice of spectral index is not too important. The source region was set to be the 90\% encircled energy radius of the PSF at $1\keV$ and the background region was an annulus centred on the same coordinates as the source region, with the inner radius equal to the source radius, and the outer radius five times greater than the inner radius. The \texttt{psfmethod} option in \texttt{srcflux} was set to \texttt{quick}, which uses the radius of the source circle to obtain the PSF fraction in the specified energy band, and assumes that the background region contains 0\% of the source flux, so the effect of any source flux that falls in the background region is neglected. The absorbing column, N\textsubscript{H}, was fixed at the Galactic value \citep{Kalberla:2005a}: $\approx$2 - 2.5$\times10^{20}$cm$^{-2}$ for all clusters except XLSSC 634 which had N\textsubscript{H}$\approx$1.5$\times10^{20}$cm$^{-2}$). All of the \texttt{wavdetect} detected point sources had ``good measurements'' from \texttt{srcflux}, except for XLSSC 072 which had ``mode of zero'' for its flux measurement so we report this as a $1\sigma$ upper limit.The fluxes are reported in column 6 in Table \ref{table:resultssummary}.

For those clusters that had no point sources detected within 60$''$ of the cluster centre, we determined an upper limit on the flux of any undetected point source. For each cluster we simulated an image of a point source, using the \emph{Chandra} PSF at the detector position of the cluster centre, and normalised to a particular point source flux. Poisson noise was added and the point source was added to the original \emph{Chandra} image at the cluster centre. We then applied the same detection method used on the original data and recorded whether the simulated point source was detected. This process was repeated for 100 realisations of the Poisson noise for a given point source flux. The source flux was then varied until the simulated source was detected in 68\% of the realisations, and the corresponding flux was defined as the 1$\sigma$ upper limit on the flux of an undetected point source. This value is reported in column 6 of Table \ref{table:resultssummary}. The upper limits are driven by the Poisson noise on the low number of counts expected from the faint point source and hence can be significantly larger than the measured flux for detected point sources in comparable observations. 

To estimate the possible contribution of point sources to the cluster flux measured with \emph{XMM}, we compute the AGN contamination fraction. The AGN contamination fraction is the contribution of the combined flux from all of the point sources detected by \emph{Chandra} (or upper limits for those clusters with no point sources detected) within 60$''$ of the cluster centre (that were not detected by XXL and excluded from the XXL flux calculation) as a fraction of F$_{60}$ (see column 4, 6 and 7 in Table \ref{table:resultssummary}). These cluster fluxes are updated compared to those from \citet{Willis:2013a}, and calculated using the updated version of the XXL analysis pipeline. Figures \ref{fig:threeimagesC1}, \ref{fig:threeimagesC2} and \ref{fig:threeimagesC3} show images of the clusters in the sample, and indicate the positions of point sources that were detected by XXL and/or by the \emph{Chandra} follow-up observations. Those detected by XXL were already excluded from the F$_{60}$ values and so do not contribute to the AGN contamination fractions calculated here. As mentioned above, the contamination was calculated as the combined point source flux (or the upper limit in the case of clean clusters) of those point sources not previously resolved by XXL as a fraction of the cluster flux. Therefore, a cluster with a contamination$\gta 1$ can be thought of as being a misclassified point source(s). Lower, but non-zero, values suggest that the XXL flux comes from a blend of cluster and point source emission.

\subsection{Calculating Cluster Fluxes from the \emph{Chandra} data} \label{subsection.clusterflux}

The \emph{Chandra} snapshot observations were optimised to detect significant point source contamination in the XXL clusters, and are not expected to be deep enough to measure detailed ICM properties. Nonetheless we attempted to place constraints on the ICM flux from the \emph{Chandra}  data. All of the point sources in the image were masked using a circle with a radius necessary to include 90\% of the flux at 1$\keV$, and the flux from each cluster was estimated using \texttt{srcflux}. A 60$''$ radius circle was used as the source region (consistent with the XXL flux measurements), and the background region used was an annulus with inner and outer radii of
120$''$ and 180$''$ respectively, as measured from the cluster centre. In some cases this background
region went off chip and this was accounted for. An absorbed APEC thermal plasma model \citep{Smith:2001a} was used to model the cluster flux. The absorption was set at the Galactic value \citep{Kalberla:2005a}, the metal abundance set to 0.3 solar, and the plasma temperature to 3.5 keV (typical of high redshift XXL clusters, XXL Paper \Rmnum{20}). The redshifts used are in Table \ref{table:samplesummary}. If the $3\sigma$ lower bound on the PDF of the flux in this region was non-zero, then we treated this as a definite detection of ICM emission with \emph{Chandra}. This was the case for five clusters. In 11 other cases, an ICM flux measurement was still possible, but the $3\sigma$ lower bound extended to zero flux. In the remaining cases the mode of the posterior distribution for the flux was zero, so only an upper limit was measured.

The effect of masking the point sources means some cluster emission is also lost from the masked region. The effect of this will be greatest for off-axis sources where the PSF and therefore the mask size is greatest. 3XLSS J021825.9-045947 has the largest PSF at cluster centre of all observations where a point source is detected in the 60$''$ cluster region (see Figure \ref{fig:threeimagesC1}). The masked region accounts for $\sim$0.5\% of the cluster area in the 60$''$ region. Modelling the cluster emission as a beta-model \citep{Cavaliere:1976a} with $\beta=0.66$ and a core radius of 175 kpc and assuming that the point source mask is at cluster centre (as this will maximise the amount of presumptive ICM flux lost) it is found that $\sim$2.5\% of the total cluster emission from the 60$''$ region is masked. Thus we can ignore this effect as the difference is much smaller than our 1$\sigma$ errors on the cluster fluxes (see Table \ref{table:resultssummary}).

\section{Notes on Individual Clusters} \label{section.indivnotes}

In this section we note any instances where we departed from the analysis described in Section 3 and other points of interest. In all cases, when PSF sizes are reported, we give the 90\% encircled energy radius at 1 keV.

For each cluster/cluster candidate below we give the name, \emph{Chandra} ObsID, XXL class, and categorise its level of AGN contamination based on all of the data available. CC indicates a ``clean cluster'' with a low level of AGN contamination; PC indicates a cluster that is ``partially contaminated'' from the point sources previously unresolved in XXL; FC indicates a ``fully contaminated'' cluster (i.e. most likely a point source - or multiple point sources - that was misclassified as extended). This information is also given in column 8 in Table \ref{table:resultssummary}.

\bigbreak
    \indent 3XLSS J021825.9-045947 / ObsID 17306 / C1 / FC - This cluster fell 2.8$'$ off-axis in an archived observation, where the PSF is 4.09$''$ compared with 0.83$''$ on-axis. A source was detected at the cluster centre but due to the larger PSF at the source position it is not clear whether this is a genuine point source or a detection of extended emission.   
However, the X-ray source is coincident with a radio source and an unresolved optical source so we conclude it is likely to
be a radio-loud quasar, and treat it as a point source. 
In addition, our \texttt{dmstat} search method identified a potential point source that was undetected by \texttt{wavdetect}, $\sim5''$ from the source that was detected at the cluster centre. 
From the optical data, there is a likely optical counterpart to this possible X-ray source that appears slightly extended in nature so is likely to be a galaxy. We thus conclude that this source (if real) is likely to be an AGN in that galaxy rather than a detection of the ICM. We do not include this undetected point source when calculating the cluster contamination, however if we were to include it the AGN contamination fraction would rise from 0.67 to 0.90. In either case it appears likely that the XXL detection is a misclassified AGN or pair of AGN and not a genuine extended source.
 \\   \indent XLSSC 122 / ObsID 18263 / C1 / CC - This cluster is at z $=$ 1.99 (based on results in \citealp[][hereafter XXL Paper \Rmnum{17},]{Mantz:2017a} using X-ray spectroscopy) and is the most distant cluster discovered by XXL to date (see XXL Paper \Rmnum{20}). It has a Sunyaev-Zel'dovich effect detection (XXL Paper \Rmnum{5}) and deep \emph{XMM} follow up (XXL Paper \Rmnum{17}). \texttt{wavdetect} found no point sources in the larger 60$''$ circular region around the cluster centre, and inspecting the image visually confirms this. We therefore computed an upper limit for contamination as described in Section \ref{subsection.ptsrcdetnandflux}. We first reported a 3$\sigma$ upper limit on the flux contamination of 8\% in
XXL paper \Rmnum{17}. Using the same \emph{Chandra} data, we here place a 1$\sigma$ upper limit
of 18\% on the flux of any undetected point source. This weaker constraint is due to the more rigorous and conservative definition of an upper limit in the current work (see Section \ref{subsection.ptsrcdetnandflux})
 \\   \indent 3XLSS J021320.3-053411 / ObsID 20535 / C2 / FC - This cluster has one point source detected in the 60$''$ cluster region by \texttt{wavdetect}. In addition, our \texttt{dmstat} search method identified a potential point source that was undetected by \texttt{wavdetect}, at 33.345, -5.56. 
There is no optical or radio counterpart for this X-ray source, and we do not include this source when calculating the cluster contamination; however its flux is 0.02$\pm$0.02 $\times$ 10$^{-14}$ erg s$^{-1}$ cm$^{-2}$ and if we were to include it the AGN contamination fraction would rise from 1.2 to 1.4. In either case it appears likely that the XXL detection is a misclassified AGN or pair of AGN and not a genuine extended source.
  \\   \indent XLSSC 203 / ObsID 17304 / C2 / FC/PC - This cluster fell 2.9$'$ off-axis in an archived observation, where the PSF is 4.59$''$ compared with 0.83$''$ on-axis. A point source was detected close to cluster centre, and upon visual inspection of the image it is clear that this is genuinely a point source (and not extended emission). The flux of this point source is about half of the XXL cluster flux, but the fluxes agree within the measurement errors, so this cluster could be partially or fully contaminated. 
 \\   \indent XLSSC 634 / ObsID 11741 / C2 / CC - This cluster fell 1.4$'$ off-axis in an archived
observation, where the PSF is 1.75$''$ compared with 0.83$''$ on-axis. A
source was detected at the cluster centre but due to the larger PSF it
is not clear whether this is a genuine point source or a detection of
extended emission. We do not find any radio or optical counterparts to this source, but conservatively treat it as point source emission for the analysis. However, if we were to treat it as ICM emission then the AGN contamination fraction would drop from 0.10 to 0.05. 
   \\   \indent 3XLSS J022005.5-050826 / ObsID 13374 / C2 / FC - For this cluster, the XXL F$_{60}$ value (see Table \ref{table:resultssummary}) has a large error, and the total flux from the 4 point sources detected in the 60$''$ cluster region is consistent with a partially contaminated cluster and also consistent with F$_{60}$ coming solely from AGN emission. However, when we mask all point sources and measure the \emph{Chandra} cluster flux (see Section \ref{subsection.clusterfluxes}), we find the cluster flux to be zero, with a low upper limit, and thus we conclude that most likely there is no cluster emission from 3XLSS J022005.5-050826, and it is multiple AGN misclassified as extended ICM emission.
  \\   \indent XLSSC 046 / ObsID 18259 / C3 / CC - This is a genuine cluster \citep{Bremer:2006a}, with an overdensity of optical and IR galaxies, but is compact, leading to its re-classification from a C2 in a previous pipeline version \citep{Willis:2013a} to a C3 with the current XXL pipeline. We did not detect any point sources in the 60$''$ cluster region with our \emph{Chandra} data.
    \\   \indent 3XLSS J022351.3-041841 / ObsID 6390 / C3 / FC/PC/CC - This cluster fell 3.7$'$ off-axis in an archived
observation, where the PSF is 6.80$''$ compared with 0.83$''$ on-axis. The centre of the cluster falls mostly on-chip, but part of the cluster emission falls off-chip. No point sources were detected in the available cluster region, so an upper limit was computed following the normal method. 
 \\   \indent 3XLSS J022812.3-043836 / ObsID 18261 / C3 / FC/PC - 
 \texttt{wavdetect} detects a point source previously detected by XXL within 60"
of the cluster centre, and for this point source the position of the centre of the ellipse enclosing the source region as detected by \texttt{wavdetect} is slightly offset from the peak pixel position when visually inspecting the image. We therefore computed the source flux at the position of the peak pixel rather than the \texttt{wavdetect} source position. When masking the point sources for the cluster flux calculation we increased the point source mask size by 1.5$''$ to ensure all of the point source emission was masked. The point source flux is reported in Table \ref{table:indivptsrcfluxessummary}, but the point source is not included in the AGN contamination fraction as it was previously detected by XXL.
 \\   \indent 3XLSS J022554.3-045059 / ObsID 18264 / C3 / FC - \texttt{wavdetect} detects three point sources within 60"
of the cluster centre. For one of the point sources, the position of the centre of the ellipse enclosing the source region as detected by \texttt{wavdetect} is slightly off from the peak pixel position when visually inspecting the image. We treated this as for  3XLSS J022812.3-043836. 


\section{Discussion} \label{section.results}

\begin{figure}
	\includegraphics[width = 88mm]{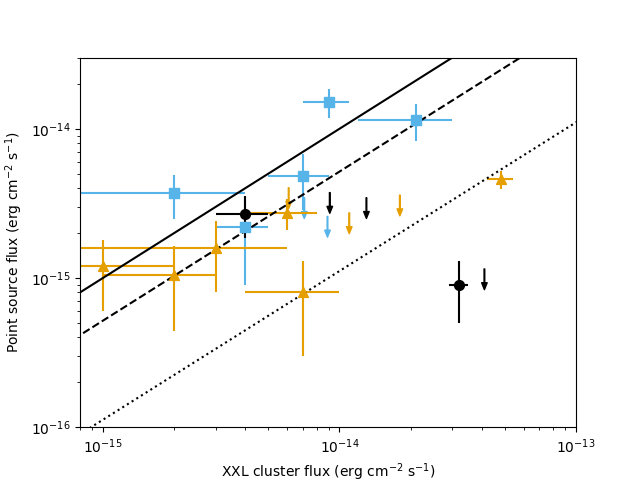}
	\caption[]{We show the total \emph{Chandra} flux for point sources within 60" of the cluster centre versus the \emph{XMM} cluster flux. C1 clusters are black circles, C2s are yellow triangles and C3s are blue squares. Arrows indicate clusters that only have a 1$\sigma$  upper limit for their point source flux (column 6 in Table \ref{table:resultssummary}) - the tip of the arrow denotes the upper limit. The solid straight line is a line of equality showing locus of 100\% AGN contamination and the dashed and dotted lines are lines of equality showing the the locus of 50\% and 10\% AGN contamination, respectively. 1$\sigma$ errors are shown.}
	\label{fig:conts}
\end{figure}

\begin{figure}
	\includegraphics[width = 90mm]{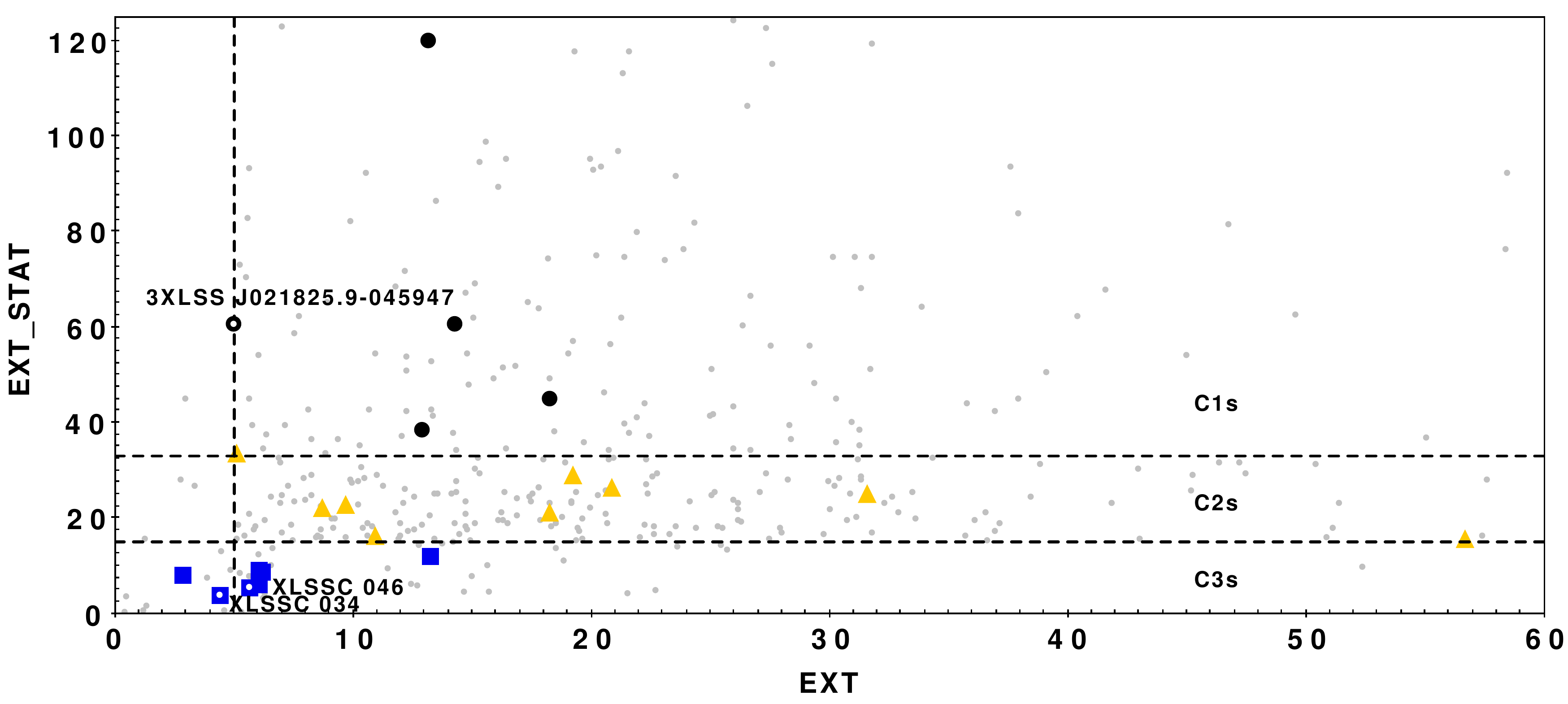} 
	\caption[]{We show the \textsc{ext} - \textsc{ext\_stat} parameter space for the C1s, C2s and C3s in our sample (larger black circles, yellow triangles and blue squares respectively). We also show a representative sample of C1, C2 and C3 XXL clusters at 0$<$z$<$1 for illustration (smaller grey circles). The C1/C2/C3 boundaries are explained in Section \ref{section.sample}. The three larger circles/squares with the hollow centres are those with labels on the plot. 
	}
	\label{fig:extent}
\end{figure}


\begin{table*}
\begin{center}


\scalebox{0.70}{
\begin{tabular}{lcccccccccc}
  \hline
  XXLID & Class & $z$ &  F$_{60}$  & No. of point &  \emph{Chandra} point source flux  & AGN contamination & Final & \emph{Chandra} cluster flux\\
       &  XXL    & &  ($10^{-14}\flux$) & sources & ($10^{-14}\flux$) & fraction & assessment & ($10^{-14}\flux$)\\
  \hline
XLSSC 072$^\star$  & C1 & 1.00  & 4.1$\pm$0.4 & 1 & $<$0.08 & $<$0.02 & CC & $3.41^{+0.85}_{-0.82}$ \\
XLSSC 029$^\star$  & C1 & 1.05  &  3.2$\pm$0.3 & 2 & 0.09$\pm$0.04 & 0.03 & CC & $3.63^{+0.30}_{-0.29}$\\
XLSSC 005$^\star$  & C1 & 1.06  &  0.9$\pm$0.2 & 0 & $<$0.26 & $<$0.29 & CC & $1.19^{+0.69}_{-0.67}$\\
3XLSS J021825.9-045947 & C1 &  1.13  &  0.4$\pm$0.1 & 1 & 0.27$\pm$0.09 & 0.67$^\dagger$ & FC &$0.32^{+0.25}_{-0.23}$ \\
XLSSC 122$^\star$   & C1 & 1.99  &  1.3$\pm$0.3 & 0 & $<$0.24 & $<$0.18 & CC & $1.98^{+0.79}_{-0.77}$\\
\hline
XLSSC 048$^\star$   & C2 & 1.01  &  1.1$\pm$0.3 & 0 & $<$0.19 & $<$0.17 & CC & $0.85^{+0.67}_{-0.63}$\\
XLSSC 073$^\star$   & C2 & 1.03  &  0.7$\pm$0.3 & 1 & 0.08$\pm$0.05 & 0.11 & CC & $0.46^{+0.41}_{-0.37}$\\
3XLSS J022755.7-043119   & C2 & 1.05  &  0.3$\pm$0.3 & 2 & 0.16$\pm$0.05 & 0.53 & FC &  $<$0.37\\
3XLSS J021320.3-053411   & C2 & 1.08  &  0.1$^{+0.2}_{-0.1}$ & 1 & 0.12$\pm$0.06 & 1.2 & FC & $0.35^{+0.32}_{-0.13}$\\
XLSSC 203    & C2 & 1.08  &  0.2$\pm$0.1 & 1 & 0.10$\pm$0.06 &  0.50 & PC & $0.54^{+0.29}_{-0.29}$\\
XLSSC 634     & C2 & 1.08  &  4.8$\pm$0.6 & 3 & 0.46$\pm$0.09 & 0.10$^\dagger$ & CC & $5.62^{+0.35}_{-0.35}$\\
3XLSS J021325.0-042000    & C2 & 1.20  &  1.5$\pm$0.5 & 0 & $<$0.25 & $<$0.17 & CC & $1.75^{+0.87}_{-0.83}$\\
3XLSS J022005.5-050826$^\star$   & C2 & 1.65  &  0.6$\pm$0.2 & 4 & 0.27$\pm$0.08 & 0.45 & FC &  $<$0.09\\
3XLSS J022418.4-043956$^\star$   & C2 & 1.67 & 0.6$\pm$0.2 & 0 & $<$0.28 & $<$0.47 & CC & $<$0.47\\
\hline
XLSSC 034     & C3 & 1.04  &  2.1$\pm$0.9 & 1 & 0.15$\pm$0.13 & 0.07 & CC & $2.52^{+0.88}_{-0.83}$\\
3XLSS J022059.0-043922$^\star$   & C3 & 1.11  &  0.9$\pm$0.3 & 1 & 1.52$\pm$0.34 & 1.7 & FC & $0.22^{+0.65}_{-0.22}$\\
XLSSC 046$^\star$   & C3 & 1.22  &  0.7$\pm$0.2 & 0 & $<$0.24 & $<$0.34 & CC & $0.99^{+0.49}_{-0.47}$\\
3XLSS J022351.3-041841    & C3 & 1.27  &  0.9$\pm$0.2 & 0 & $<$0.18 & $<$0.20$^\dagger$ & FC/PC/CC & $0.17^{+0.36}_{-0.17}$\\
3XLSS J021700.4-034746$^\star$   & C3 & 1.54  &  0.7$\pm$0.2 & 2 & 0.48$\pm$0.20 & 0.69 & FC & $<$0.45\\
3XLSS J022812.3-043836$^\star$  & C3 & 1.67  &  0.4$\pm$0.1 & 1 & 0.22$\pm$0.13 & 0.55 & FC/PC &   $0.54^{+0.51}_{-0.48}$\\
3XLSS J022554.3-045059$^\star$  & C3 & 2.24  &  0.2$\pm$0.2 & 2 & 0.37$\pm$0.12 & 1.9 & FC & $0.07^{+0.47}_{-0.07}$\\
   \hline
\end{tabular}
}
\caption{\label{table:resultssummary}Summary of point source detection and cluster contamination from the \textit{Chandra} data. The \emph{Chandra} cluster flux measurement is also shown. Column 4 is the XXL cluster flux. Column 5 gives the number of point sources detected by \texttt{wavdetect} within a 60$''$ radius region around the cluster centre that were not  previously detected by XXL
. Column 6 gives the total flux of all of the point sources detected by \texttt{wavdetect} within a 60$''$ region around the cluster centre that weren't detected by XXL,
with the 1$\sigma$ lower and upper limits are given as error. All fluxes are in the 0.5 - 2 keV energy band. Column 7 gives the fraction of F$_{60}$ resolved into point sources by Chandra, as described in Section \ref{subsection.ptsrcdetnandflux}. Column 8 gives our assessment of the cluster. Column 9 is the cluster flux as calculated from \emph{Chandra} data after point source removal (described in Section \ref{subsection.clusterflux}) with 1$\sigma$ errors. XXLIDs marked with a $\star$ appear in \citet{Willis:2013a} and are therefore part of the XMM-LSS subset of clusters. AGN contamination fractions marked with a $\dagger$ have possible additional contamination from potential point sources that did not meet our detection threshold (except for XLSSC 634 which has a potentially lower AGN contamination fraction than stated in this table), and  contamination values that include these sources are given in Section \ref{section.indivnotes}. Individual point source fluxes and positions are given in Table \ref{table:indivptsrcfluxessummary} in Appendix A}
\end{center}
\end{table*}

\subsection{Cluster Contaminations} \label{subsection.clustercontmns}

\begin{figure}
	\includegraphics[width = 82mm]{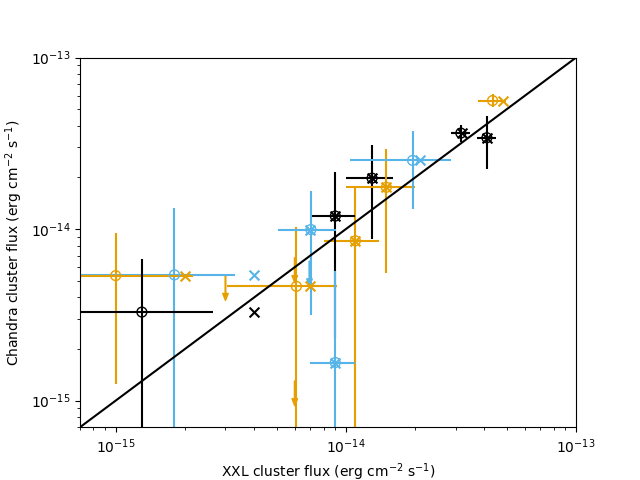}
	\caption[]{We show the \emph{Chandra} cluster flux versus the XXL cluster flux, F$_{60}$. C1 clusters are black circles/crosses/arrows, C2s are yellow circles/crosses/arrows and C3s are blue circles/crosses/arrows. The crosses are F$_{60}$ as listed in Table \ref{table:resultssummary} column 4 (i.e. the original flux, not excluding the point sources detected by Chandra). The circles are the F$_{60}$ minus the flux from any point source detected in the \emph{Chandra} data that was not previously resolved by XXL data (listed in Table \ref{table:resultssummary} column 6). The solid line is a line of equality. The arrows indicate upper limits on the \emph{Chandra} cluster flux - the tip of the arrow denotes the upper limit 
 and are plotted against the point source corrected XXL flux. 3XLSS J022059.0-043922 and 3XLSS J022554.3-045059 are not shown on the plot as the \emph{Chandra} point source flux is greater than F$_{60}$.}
	\label{fig:cluster_fluxes_comparison}
\end{figure}

We report the point source detections, fluxes and cluster contaminations in Table \ref{table:resultssummary}. Individual point source flux measurements for each cluster can be found in Table \ref{table:indivptsrcfluxessummary}. We plot the point source flux against the cluster flux to show the contamination levels in Figure \ref{fig:conts}. 


Our results provide an important validation of the performance of the XXL cluster detection pipeline in classifying distant clusters.  Four out of five of the C1 clusters are genuine uncontaminated clusters. Only the C1 3XLSS J021825.9-045947 is contaminated by AGN to a significant level (67\% contamination, or 90\% if we include the second undetected point source as discussed in Section \ref{section.indivnotes}). The C1 class is expected to be free from strongly contaminated clusters or misclassified AGN, but in this case the source was
precisely at the threshold value in extension required for classification as a cluster. This is illustrated in Figure \ref{fig:extent} which shows the clusters and cluster candidates in the \textsc{ext} - \textsc{ext\_stat} parameter space. Furthermore, this cluster was detected 6$'$ off-axis in the \emph{XMM} observation making extent measurements more challenging due to the increased asymmetry of the PSF. This appears to be a rare case of a false-positive C1 cluster at the classification threshold. 

The C2 class shows a higher level of contamination than the C1 class, as expected - five clusters have no significant point source contamination (we include XLSSC 634 here, as, despite having five point sources detected in the 60$''$ cluster region, three of which were not detected by XXL, their contribution to F$_{60}$ is very low) and the other four (3XLSS J022005.5-050826, XLSSC 203, 3XLSS J022755.7-043119 and 3XLSS J021320.3-053411) are either a blend of cluster and AGN emission or misclassified AGN. Our \emph{Chandra} cluster flux measurement suggests that 3XLSS J022005.5-050826 is not a genuine cluster, as the 1$\sigma$ upper limit for the cluster flux is low (see column 9 of Table \ref{table:resultssummary}). The results from our C2 clusters are consistent with the $<$ 50\% contamination expected in the C2 sample. The results from our C2 clusters are consistent with the $<$ 50\% contamination that is expected in the C2 sample, and demonstrate that the XXL detection pipeline is capable of detecting extended sources even in the presence of relatively bright point sources.

Looking at the 14 C1 and C2 clusters together, nine have either no newly resolved point sources, or have new \emph{Chandra}-detected sources that do not contribute significantly to the ICM flux (i.e. $>$ 15\%). A further cluster, XLSSC 203, is more strongly contaminated (at the 50\% level) but the \emph{Chandra} measurement of the ICM flux from this system supports the conclusion that it comprises a blend of ICM and point source flux. The clusters form a useful sample that can be targeted for deeper follow up observations to probe ICM properties at z $>$ 1 with good limits on the systematics from point source contamination. The legacy value of this should not be underestimated - there is no approved mission that will replace \emph{Chandra}'s imaging capabilities.


We can compare the updated pipeline (XXL Paper \Rmnum{20}) directly to that used by \citet{Willis:2013a}. 
If we define a ``clean'' cluster 
as having an AGN contamination fraction less than 0.15 for cases where \texttt{wavdetect} detects a point source within 60$''$ of the cluster centre, or a cluster that has no point sources detected by \texttt{wavdetect} in this region, we can see that the updated pipeline is more conservative.
There is an improvement for the C2 class with the updated pipeline, giving us a more robust sample with 5/9 C2s clean, compared with 2/7 using the \citet{Willis:2013a} classes.

The 7 C3 candidates were optically selected and associated with XXL sources that do not meet the criteria for the C1 or C2 classes. As would be expected, this sample is less pure than the other classes, but two of the C3s are unambiguous high-z clusters, on the basis of low contamination fractions, supporting optical data and robust ICM detections in XXL and \emph{Chandra} data. XLSSC 034 has a low level of contamination, and XLSSC 046 is a genuine cluster that was studied in detail by \citet{Bremer:2006a}.

These C3 clusters do not have a well-defined selection function, but still present interesting targets for further study. Additional such clusters could be recovered by studying the optical/IR data for sources in the same \textsc{ext} - \textsc{ext\_stat} parameter space (see Section \ref{section.sample}) as XLSSC 046. The location of XLSSC 034 and XLSSC 046 in the \textsc{ext} - \textsc{ext\_stat} parameter space is shown in Figure \ref{fig:extent}. 

We note that the existence of clusters like the C3s that fail to meet
the main survey selection criteria, and the presence of AGN
contamination in the C1/C2 sample, does not represent a problem for the
XXL selection function. The results of these snapshots validate the current modelling of the survey selection function,
and provide useful additional input for its further refinement and
testing by hydrodynamical simulations.


 \citet{Galametz:2009a} studied X-ray selected AGN in galaxy clusters that were selected in the infrared. If we apply the same selection to the AGN detected in our \emph{Chandra} observations, we would not detect any AGN in the inner 0.25 Mpc of our C1 and C2 clusters in the redshift range $1 < z < 1.5$. This is not inconsistent with the results from \citet{Galametz:2009a}, since based on their detection rate, we would expect $\sim1$ AGN to be detected in the C1 and C2 cluster sample. Our results show that the effect of selecting clusters in the X-ray band does not strongly bias our sample towards clusters containing X-ray bright AGN. 

A potentially important issue that has not yet been addressed is that of the variability of AGN. The \emph{XMM} data used in the XMM-XXL survey were mostly taken years before the \emph{Chandra} follow-up (this is true for at least the non-archival data that are the majority of our data). The typical variability in flux of AGN on this timescale is $\sim$50\% \citep[][in prep.]{Maughan:2018a}. Therefore, any cluster found to have a low (or undetectable) level of AGN contamination is unlikely to have been $\gta$30\% contaminated at the epoch of the XXL observation (or indeed at the epoch of any future, deeper observations). 

%
%

\subsection{ICM Fluxes} \label{subsection.clusterfluxes}

The cluster fluxes calculated from our \emph{Chandra} data are shown in Table \ref{table:resultssummary} and are compared with the XXL fluxes in Figure \ref{fig:cluster_fluxes_comparison}. For four of the clusters (XLSSC 072, XLSSC 029, XLSSC 634, XLSSC 034) the $3\sigma$ lower limit on the flux is greater than zero. The rest of the C1 and C2 clusters have $1\sigma$ lower limit greater than zero, except for 3XLSS J022755.7-043119, 3XLSS J022005.5-050826 and 3XLSS J022418.4-043956. These three clusters have upper limits that are consistent with the XXL flux (accounting for the unresolved AGN in the F$_{60}$ measurement). In summary, after accounting for unresolved AGN in the XXL measurements and the measurement uncertainties, all of the cluster fluxes calculated from our \emph{Chandra} data are consistent with those from XXL.


In some cases the \emph{Chandra} cluster flux is non-zero, even when we believe there is only AGN emission and no cluster emission (3XLSS J021825.9-045947, 3XLSS J022059.0-043922, 3XLSS J021320.3-053411). In these cases, the \emph{Chandra} ICM fluxes are not significantly different
from zero and we interpret the signals as noise fluctuations rather
than ICM detections.


\section{Conclusion} \label{section.conclusion}
We have analysed \emph{Chandra} data for 21 clusters and cluster candidates that appear in the XMM-XXL survey catalogue in order to determine the extent of any contamination by unresolved point sources. Our main results are as follows:
\begin{itemize}
\item In the 14 C1 and C2 clusters which form a complete sample with a defined selection function, we find that the majority have little or no contamination of their ICM fluxes by AGN. One C1 source appears to be an AGN that was misclassified as extended, but this source was detected at the extension parameter threshold, so represents a rare interloper rather than any broad problem in the classification scheme. Three or four of the nine C2 clusters are either AGN that were misclassified as extended sources, or else have ICM emission that is strongly contaminated by AGN emission. Overall these results agree well with the calibration of the XXL selection function and serve to validate its description of these distant cluster samples. We remind the reader that these conclusions were derived for distant clusters where the angular size of a cluster might be a similar size to the \emph{XMM} PSF; therefore, our conclusions should not be extrapolated to the lower redshift XXL clusters.

\item With this \emph{Chandra} follow-up, we have defined a complete sample of ten $z > 1$ clusters (those marked CC in column 8 Table \ref{table:resultssummary} and XLSSC 203) for further study. This comprises all secure C1 and C2 clusters that show evidence for X-ray emission originating from the ICM (in addition to any contaminating AGN if they are detected).

\item Of the seven C3 optically selected cluster candidates with X-ray counterparts that did not meet the C1 or C2 selection criteria, we consider two (XLSSC 034 and XLSSC 046) to be genuine clusters with low levels of AGN contamination. A third, 3XLSS J0222351.3-041841 may also be a genuine cluster with low contamination, but this is unclear due to the cluster region being only partially covered by \emph{Chandra}. The remaining four sources are either AGN or clusters with high levels of AGN contamination.

\item We measured the ICM flux with \emph{Chandra}, recording upper limits in three cases. For all clusters, the \emph{Chandra} ICM flux was consistent with that measured by \emph{XMM} once the \emph{XMM} flux was corrected for unresolved point sources.

\item The number of AGN per cluster for this X-ray selected sample was found to be lower, but consistent with, that of clusters selected in the infra-red, indicating the X-ray selection with the XXL pipeline does not lead to a bias towards clusters with associated X-ray bright AGN.
\end{itemize}

We have demonstrated the utility of \emph{Chandra} snapshots to test for AGN in or projected onto clusters detected in surveys with poorer resolution, for example the upcoming \emph{eROSITA}  survey, which has a HEW of 28$''$ average over the entire field of view \citep{Merloni:2012a,Pillepich:2012a}. \emph{Chandra} snapshots can be used to decontaminate \emph{eROSITA} high-z candidate clusters using methods similar to those presented in this paper. 

\section{Acknowledgements}
XXL is an international project based around an \emph{XMM} Very Large Programme surveying two 25 deg$^2$ extragalactic fields at a depth of $\sim$ 6 x 10$^{-15}$ erg cm$^{-2}$ s$^{-1}$ in the 0.5- 2 keV band for point-like sources. The XXL website is \url{http://irfu.cea.fr/xxl}. Multi-band information and spectroscopic follow-up of the X-ray sources are obtained through a number of survey programmes, summarised at \url{http://xxlmultiwave.pbworks.com}.
We thank Adam Mantz and Mauro Sereno for their useful comments on this work. 
FP and MERC acknowledge support by the German
Aerospace Agency (DLR) with funds from the Ministry of
Economy and Technology (BMWi) through grant 50 OR 1514. The Saclay group acknowledges long-term support from the Centre National d'Etudes Spatiales (CNES).
This paper has also made use of observations obtained with the Canada-France-Hawaii Telescope (CFHT) which is operated by the National Research Council (NRC) of Canada, the Institut National des Sciences de l\textsc{'}Univers of the Centre National de la Recherche Scientifique(CNRS) of France, and the University of Hawaii. This paper has also made use of the NASA/IPAC Extragalactic Database (NED) which is operated by the Jet
Propulsion Laboratory, California Institute of Technology, under contract with
the National Aeronautics and Space Administration. This paper has also made use of \textsc{topcat} \citep{Taylor:2005a}.

\bibliographystyle{aa}
\bibliography{/Users/cl16977/Documents/emacs_files/chandra_snapshots/Snapshots} 

\appendix

\section{Point Source Positions and Individual Fluxes}

\begin{table*}
\begin{center}
\scalebox{0.88}{
\begin{tabular}{lcccccccccc}
  \hline
  XXLID   & Class & $z$ & RA & Dec. & Flux   & Resolved  & Separation from    \\
       &   XXL   & & (J2000) & (J2000) & ($10^{-14}\flux$) & by \emph{XMM}  & cluster centre ($''$)  \\
  \hline
XLSSC 072$^\star$   & C1 & 1.00 & 33.852 & -3.726 &  $<$0.08  & No & 8 \\
XLSSC 029$^\star$     & C1 & 1.05 & 36.002 & -4.225 &  0.04$^{+0.03}_{-0.02}$  & No & 52 \\
  &   &  & 36.012 & -4.229 & 0.05$^{+0.03}_{-0.02}$ & No & 23 \\
XLSSC 005$^\star$   & C1 & 1.06 & - & - &  -  & -  & - \\
3XLSS J021825.9-045947    & C1 &  1.13 & 34.609 & -4.996 & 0.27$^{+0.09}_{-0.08}$  & No & 5 \\
XLSSC 122$^\star$   & C1 & 1.99 & - & - &  -  & - & - \\
\hline
XLSSC 048$^\star$   & C2 & 1.01 & - & - &  -  & - & - \\
XLSSC 073$^\star$   & C2 & 1.03 & 33.749 & -3.515 &  0.08$^{+0.08}_{-0.05}$ & No & 37  \\
  & &   & 33.737 & -3.519 &  0.37$^{+0.14}_{-0.12}$ & 3XLSS J021456.8-033108 & 53 \\
3XLSS J022755.7-043119   & C2 & 1.05 &36.972 & -4.516 &  0.05$^{+0.05}_{-0.03}$ & No & 10  \\ 
& &   & 36.984 & -4.521 &  0.11$^{+0.06}_{-0.05}$ & No & 43  \\
& &   & 36.994 & -4.520 &  0.61$^{+0.14}_{-0.12}$ & 3XLSS J022758.7-043110 & 44 \\
3XLSS J021320.3-053411   & C2 & 1.08 & 33.331 & -3.571 &  0.12$^{+0.06}_{-0.05}$ & No & 11 \\
XLSSC 203    & C2 & 1.08 & 34.429 & -4.988 & 0.10$^{+0.07}_{-0.05}$   & No  & 4 \\
XLSSC 634    & C2 & 1.08 & 355.692 & -54.185 &  0.20$^{+0.06}_{-0.06}$ & No & 4  \\
  &  &  & 355.704 & -54.185 & 0.22$^{+0.06}_{-0.05}$  & No & 29 \\
  &  &  & 355.683 & -54.177 &  1.30$^{+0.14}_{-0.14}$ & 3XLSS J234244.2-541033 & 34 \\
  &  &  & 355.687 & -54.175 &  0.09$^{+0.05}_{-0.03}$ & 3XLSS J234244.2-541033 & 36 \\
  &  &  & 355.712 & -54.176 &  0.04$^{+0.02}_{-0.02}$  & No & 54 \\
3XLSS J021325.0-042000    & C2 & 1.20 & - & - & -  & - & - \\
3XLSS J022005.5-050826$^\star$   & C2 & 1.65 & 35.022 & -5.140 &  0.15$^{+0.05}_{-0.03}$ & No & 5 \\
 &  &   & 35.021 & -5.139 & 0.01$^{+0.02}_{-0.01}$  & No & 10 \\
 &  &  & 35.030 & -5.137 & 0.07$^{+0.05}_{-0.04}$ & No & 29  \\
 &  &  & 35.014 & -5.134 & 0.04$^{+0.03}_{-0.02}$  & No & 41 \\
3XLSS J022418.4-043956$^\star$  & C2 & 1.67 & - & - & -  & - & - \\
\hline
XLSSC 034  &  C3 & 1.04 & 35.372 & -4.093 & 1.00$^{+0.32}_{-0.26}$ & 3XLSS J022129.1-040534 & 22  \\
 &  &  & 35.375 & -4.111 & 0.15$^{+0.16}_{-0.10}$ & No & 45 \\
3XLSS J022059.0-043922$^\star$   & C3 & 1.11 & 35.247 & -4.656 & 1.52$^{+0.37}_{-0.32}$  & No & 5 \\
XLSSC 046$^\star$   & C3 & 1.22 & - & - &  -  & - & - \\
3XLSS J022351.3-041841      & C3 & 1.27 & - & - &  - & - & - \\
3XLSS J021700.4-034746$^\star$   & C3 & 1.54 & 34.253 & -3.795 & 0.41$^{+0.20}_{-0.15}$  & No & 8  \\
&  & &   34.258 & -3.784 & 0.07$^{+0.11}_{-0.06}$  & No & 50 \\
3XLSS J022812.3-043836$^\star$   & C3 & 1.67 & 37.051 & -4.651 & 0.10$^{+0.04}_{-0.04}$  & 3XLSS J022812.2-043906 & 25 \\
&  &  &   37.045 & -4.648 & 0.22$^{+0.15}_{-0.11}$  & No & 26 \\
3XLSS J022554.3-045059$^\star$   & C3 & 2.24 & 36.477 & -4.851 & 0.13$^{+0.08}_{-0.06}$  & No  & 5 \\
 &  &  &   36.472 & -4.846 & 0.24$^{+0.10}_{-0.08}$  & No & 20 \\
 &  &  &   36.471 & -4.837 & 0.89$^{+0.19}_{-0.16}$ & 3XLSS J022552.8-045013 & 50 \\
    \hline
\end{tabular}
}
\caption{\label{table:indivptsrcfluxessummary} Summary of the fluxes for all point sources within 60$''$ of the cluster centre
. Column 6 is the individual point source flux as calculated from the \emph{Chandra} data with 1$\sigma$ errors. All fluxes are in the 0.5 - 2 keV energy band. XXLIDs marked with a $\star$ appear in \citet{Willis:2013a} and are therefore part of the XMM-LSS subset of clusters. 
Column 7 states whether the \emph{Chandra} detected point source was previously resolved by XXL and thus excluded from the F$_{60}$ measurements; for cases where the point source was resolved by XXL, its name as in XXL Paper \Rmnum{27} is provided. In the case of XLSSC 634, two sources were blended into one by the \emph{XMM} PSF, reported as one point source by XXL, and were masked from the F$_{60}$ calculation.}
\end{center}
\end{table*}

\end{document}

%% file: newcom.tex
\newcommand{\gta}{\,\rlap{\raise 0.4ex\hbox{$>$}}{\lower 0.6ex\hbox{$\sim$}}\,}
\newcommand{\lta}{\,\rlap{\raise 0.4ex\hbox{$<$}}{\lower 0.6ex\hbox{$\sim$}}\,}

\newcommand{\m}{\mbox{\ensuremath{\mathrm{~m}}}}

\newcommand{\s}{\mbox{\ensuremath{\mathrm{~s}}}}

\newcommand{\keV}{\mbox{\ensuremath{\mathrm{~keV}}}}
\newcommand{\erg}{\mbox{\ensuremath{\mathrm{~erg}}}}

\newcommand{\muT}{\mbox{\ensuremath{~\mathrm{\muT}}}}

\newcommand{\pcmsq}{\mbox{\ensuremath{\mathrm{~cm^{-2}}}}}

\newcommand{\ps}{\ensuremath{\mathrm{\s^{-1}}}}

\newcommand{\flux}{\ensuremath{\mathrm{\erg \ps \pcmsq}}}



%% file: chandra_snapshots_write_up_aa_for_arxiv_and_A_A_13_Aug.bbl
\begin{thebibliography}{60}
\expandafter\ifx\csname natexlab\endcsname\relax\def\natexlab#1{#1}\fi

\bibitem[{{Adami} {et~al.}(2018){Adami}, {Giles}, {Pacaud}, {Caretta},
  {Pierre}, {Eckert}, \& {Ramos-Ceja}}]{Adami:2018a}
{Adami}, C., {Giles}, P., {Pacaud}, F., {et~al.} 2018, \aap, In preparation,
  XXL Paper XX

\bibitem[{{Adami} {et~al.}(2011){Adami}, {Mazure}, {Pierre}, {Sprimont},
  {Libbrecht}, {Pacaud}, {Clerc}, {Sadibekova}, {Surdej}, {Altieri}, {Duc},
  {Galaz}, {Gueguen}, {Guennou}, {Hertling}, {Ilbert}, {Le F{\`e}vre},
  {Quintana}, {Valtchanov}, {Willis}, {Akiyama}, {Aussel}, {Chiappetti},
  {Detal}, {Garilli}, {Lebrun}, {Lef{\`e}vre}, {Maccagni}, {Melin}, {Ponman},
  {Ricci}, \& {Tresse}}]{Adami:2011a}
{Adami}, C., {Mazure}, A., {Pierre}, M., {et~al.} 2011, \aap, 526, A18

\bibitem[{{Allen} {et~al.}(2011){Allen}, {Evrard}, \& {Mantz}}]{Allen:2011a}
{Allen}, S.~W., {Evrard}, A.~E., \& {Mantz}, A.~B. 2011, \araa, 49, 409

\bibitem[{{Biffi} {et~al.}(2018){Biffi}, {Dolag}, \& {Merloni}}]{Biffi:2018a}
{Biffi}, V., {Dolag}, K., \& {Merloni}, A. 2018, ArXiv e-prints

\bibitem[{{B{\"o}hringer} {et~al.}(2004){B{\"o}hringer}, {Schuecker}, {Guzzo},
  {Collins}, {Voges}, {Cruddace}, {Ortiz-Gil}, {Chincarini}, {De Grandi},
  {Edge}, {MacGillivray}, {Neumann}, {Schindler}, \&
  {Shaver}}]{Bohringer:2004a}
{B{\"o}hringer}, H., {Schuecker}, P., {Guzzo}, L., {et~al.} 2004, \aap, 425,
  367

\bibitem[{{Branchesi} {et~al.}(2007){Branchesi}, {Gioia}, {Fanti}, \&
  {Fanti}}]{Branchesi:2007a}
{Branchesi}, M., {Gioia}, I.~M., {Fanti}, C., \& {Fanti}, R. 2007, \aap, 472,
  727

\bibitem[{{Bremer} {et~al.}(2006){Bremer}, {Valtchanov}, {Willis}, {Altieri},
  {Andreon}, {Duc}, {Fang}, {Jean}, {Lonsdale}, {Pacaud}, {Pierre}, {Shupe},
  {Surace}, \& {Waddington}}]{Bremer:2006a}
{Bremer}, M.~N., {Valtchanov}, I., {Willis}, J., {et~al.} 2006, \mnras, 371,
  1427

\bibitem[{{Butler} {et~al.}(2017){Butler}, {Huynh}, {Delhaize}, {Smol{\v
  c}i{\'c}}, {Kapi{\'n}ska}, {Milakovi{\'c}}, {Novak}, {Baran}, {O'Brien},
  {Chiappetti}, {Desai}, {Fotopoulou}, {Horellou}, {Lidman}, \&
  {Pierre}}]{Butler:2017a}
{Butler}, A., {Huynh}, M., {Delhaize}, J., {et~al.} 2017, ArXiv e-prints, XXL
  Paper XVIII

\bibitem[{{Cavaliere} \& {Fusco-Femiano}(1976)}]{Cavaliere:1976a}
{Cavaliere}, A. \& {Fusco-Femiano}, R. 1976, \aap, 49, 137

\bibitem[{{Chiappetti} {et~al.}(2018){Chiappetti}, {Fotopolou}, {Lidman}, \&
  {Faccioli}}]{Chiappetti:2018a}
{Chiappetti}, L., {Fotopolou}, S., {Lidman}, C., \& {Faccioli}, L. 2018, \aap,
  Accepted

\bibitem[{{Clerc} {et~al.}(2012){Clerc}, {Sadibekova}, {Pierre}, {Pacaud}, {Le
  F{\`e}vre}, {Adami}, {Altieri}, \& {Valtchanov}}]{Clerc:2012a}
{Clerc}, N., {Sadibekova}, T., {Pierre}, M., {et~al.} 2012, \mnras, 423, 3561

\bibitem[{{Condon} {et~al.}(1998){Condon}, {Cotton}, {Greisen}, {Yin},
  {Perley}, {Taylor}, \& {Broderick}}]{Condon:1998a}
{Condon}, J.~J., {Cotton}, W.~D., {Greisen}, E.~W., {et~al.} 1998, \aj, 115,
  1693

\bibitem[{{Desai} {et~al.}(2012){Desai}, {Armstrong}, {Mohr}, {Semler}, {Liu},
  {Bertin}, {Allam}, {Barkhouse}, {Bazin}, {Buckley-Geer}, {Cooper}, {Hansen},
  {High}, {Lin}, {Lin}, {Ngeow}, {Rest}, {Song}, {Tucker}, \&
  {Zenteno}}]{Desai:2012a}
{Desai}, S., {Armstrong}, R., {Mohr}, J.~J., {et~al.} 2012, \apj, 757, 83

\bibitem[{{Ebeling} {et~al.}(1998){Ebeling}, {Edge}, {Bohringer}, {Allen},
  {Crawford}, {Fabian}, {Voges}, \& {Huchra}}]{Ebeling:1998a}
{Ebeling}, H., {Edge}, A.~C., {Bohringer}, H., {et~al.} 1998, \mnras, 301, 881

\bibitem[{{Ebeling} {et~al.}(2010){Ebeling}, {Edge}, {Mantz}, {Barrett},
  {Henry}, {Ma}, \& {van Speybroeck}}]{Ebeling:2010a}
{Ebeling}, H., {Edge}, A.~C., {Mantz}, A., {et~al.} 2010, \mnras, 407, 83

\bibitem[{{Eisenhardt} {et~al.}(2008){Eisenhardt}, {Brodwin}, {Gonzalez},
  {Stanford}, {Stern}, {Barmby}, {Brown}, {Dawson}, {Dey}, {Doi}, {Galametz},
  {Jannuzi}, {Kochanek}, {Meyers}, {Morokuma}, \&
  {Moustakas}}]{Eisenhardt:2008a}
{Eisenhardt}, P.~R.~M., {Brodwin}, M., {Gonzalez}, A.~H., {et~al.} 2008, \apj,
  684, 905

\bibitem[{{Faccioli} {et~al.}(2018){Faccioli}, {Pacaud}, {Pierre}, {Valotti},
  {Pacaud}, {Clerc}, \& {Gastaud}}]{Faccioli:2018a}
{Faccioli}, L., {Pacaud}, F., {Pierre}, M., {et~al.} 2018, \aap, Submitted, XXL
  Paper XXIV

\bibitem[{{Fassbender} {et~al.}(2011){Fassbender}, {Nastasi}, {B{\"o}hringer},
  {{\v S}uhada}, {Santos}, {Rosati}, {Pierini}, {M{\"u}hlegger}, {Quintana},
  {Schwope}, {Lamer}, {de Hoon}, {Kohnert}, {Pratt}, \&
  {Mohr}}]{Fassbender:2011a}
{Fassbender}, R., {Nastasi}, A., {B{\"o}hringer}, H., {et~al.} 2011, \aap, 527,
  L10

\bibitem[{{Fotopoulou} {et~al.}(2016){Fotopoulou}, {Pacaud}, {Paltani},
  {Ranalli}, {Ramos-Ceja}, {Faccioli}, {Plionis}, {Adami}, {Bongiorno},
  {Brusa}, {Chiappetti}, {Desai}, {Elyiv}, {Lidman}, {Melnyk}, {Pierre},
  {Piconcelli}, {Vignali}, {Alis}, {Ardila}, {Arnouts}, {Baldry}, {Bremer},
  {Eckert}, {Guennou}, {Horellou}, {Iovino}, {Koulouridis}, {Liske},
  {Maurogordato}, {Menanteau}, {Mohr}, {Owers}, {Poggianti}, {Pompei},
  {Sadibekova}, {Stanford}, {Tuffs}, \& {Willis}}]{Fotopoulou:2016a}
{Fotopoulou}, S., {Pacaud}, F., {Paltani}, S., {et~al.} 2016, \aap, 592, A5,
  XXL Paper IV

\bibitem[{{Fruscione} {et~al.}(2006){Fruscione}, {McDowell}, {Allen},
  {Brickhouse}, {Burke}, {Davis}, {Durham}, {Elvis}, {Galle}, {Harris},
  {Huenemoerder}, {Houck}, {Ishibashi}, {Karovska}, {Nicastro}, {Noble},
  {Nowak}, {Primini}, {Siemiginowska}, {Smith}, \& {Wise}}]{Fruscione:2006a}
{Fruscione}, A., {McDowell}, J.~C., {Allen}, G.~E., {et~al.} 2006, in
  \procspie, Vol. 6270, Society of Photo-Optical Instrumentation Engineers
  (SPIE) Conference Series, 62701V

\bibitem[{{Galametz} {et~al.}(2009){Galametz}, {Stern}, {Eisenhardt},
  {Brodwin}, {Brown}, {Dey}, {Gonzalez}, {Jannuzi}, {Moustakas}, \&
  {Stanford}}]{Galametz:2009a}
{Galametz}, A., {Stern}, D., {Eisenhardt}, P.~R.~M., {et~al.} 2009, \apj, 694,
  1309

\bibitem[{{Giles} {et~al.}(2012){Giles}, {Maughan}, {Birkinshaw}, {Worrall}, \&
  {Lancaster}}]{Giles:2012a}
{Giles}, P.~A., {Maughan}, B.~J., {Birkinshaw}, M., {Worrall}, D.~M., \&
  {Lancaster}, K. 2012, \mnras, 419, 503

\bibitem[{{Giles} {et~al.}(2016){Giles}, {Maughan}, {Pacaud}, {Lieu}, {Clerc},
  {Pierre}, {Adami}, {Chiappetti}, {D{\'e}mocl{\'e}s}, {Ettori}, {Le
  F{\'e}vre}, {Ponman}, {Sadibekova}, {Smith}, {Willis}, \&
  {Ziparo}}]{Giles:2016a}
{Giles}, P.~A., {Maughan}, B.~J., {Pacaud}, F., {et~al.} 2016, \aap, 592, A3,
  XXL Paper III

\bibitem[{{Gioia} {et~al.}(1990){Gioia}, {Maccacaro}, {Schild}, {Wolter},
  {Stocke}, {Morris}, \& {Henry}}]{Gioia:1990a}
{Gioia}, I.~M., {Maccacaro}, T., {Schild}, R.~E., {et~al.} 1990, \apjs, 72, 567

\bibitem[{{Gladders} \& {Yee}(2005)}]{Gladders:2005a}
{Gladders}, M.~D. \& {Yee}, H.~K.~C. 2005, \apjs, 157, 1

\bibitem[{{Hilton} {et~al.}(2010){Hilton}, {Lloyd-Davies}, {Stanford}, {Stott},
  {Collins}, {Romer}, {Hosmer}, {Hoyle}, {Kay}, {Liddle}, {Mehrtens}, {Miller},
  {Sahl{\'e}n}, \& {Viana}}]{Hilton:2010a}
{Hilton}, M., {Lloyd-Davies}, E., {Stanford}, S.~A., {et~al.} 2010, \apj, 718,
  133

\bibitem[{{Hinshaw} {et~al.}(2013){Hinshaw}, {Larson}, {Komatsu}, {Spergel},
  {Bennett}, {Dunkley}, {Nolta}, {Halpern}, {Hill}, {Odegard}, {Page}, {Smith},
  {Weiland}, {Gold}, {Jarosik}, {Kogut}, {Limon}, {Meyer}, {Tucker}, {Wollack},
  \& {Wright}}]{Hinshaw:2013a}
{Hinshaw}, G., {Larson}, D., {Komatsu}, E., {et~al.} 2013, \apjs, 208, 19

\bibitem[{{Kalberla} {et~al.}(2005){Kalberla}, {Burton}, {Hartmann}, {Arnal},
  {Bajaja}, {Morras}, \& {P{\"o}ppel}}]{Kalberla:2005a}
{Kalberla}, P.~M.~W., {Burton}, W.~B., {Hartmann}, D., {et~al.} 2005, \aap,
  440, 775

\bibitem[{{Kim} {et~al.}(2007){Kim}, {Kim}, {Wilkes}, {Green}, {Kim},
  {Anderson}, {Barkhouse}, {Evans}, {Ivezi{\'c}}, {Karovska}, {Kashyap}, {Lee},
  {Maksym}, {Mossman}, {Silverman}, \& {Tananbaum}}]{Kim:2007a}
{Kim}, M., {Kim}, D.-W., {Wilkes}, B.~J., {et~al.} 2007, \apjs, 169, 401

\bibitem[{{Koulouridis} {et~al.}(2017){Koulouridis}, {Faccioli}, {Le Brun},
  {Plionis}, {McCarthy}, {Pierre}, {Akylas}, {Georgantopoulos}, {Paltani},
  {Lidman}, {Fotopoulou}, {Vignali}, {Pacaud}, \&
  {Ranalli}}]{Koulouridis:2017a}
{Koulouridis}, E., {Faccioli}, L., {Le Brun}, A.~M.~C., {et~al.} 2017, ArXiv
  e-prints

\bibitem[{{Koulouridis} {et~al.}(2014){Koulouridis}, {Plionis}, {Melnyk},
  {Elyiv}, {Georgantopoulos}, {Clerc}, {Surdej}, {Chiappetti}, \&
  {Pierre}}]{Koulouridis:2014a}
{Koulouridis}, E., {Plionis}, M., {Melnyk}, O., {et~al.} 2014, \aap, 567, A83

\bibitem[{{Mantz} {et~al.}(2017){Mantz}, {Abdulla}, {Allen}, {Carlstrom},
  {Logan}, {Marrone}, {Maughan}, {Willis}, {Pacaud}, \& {Pierre}}]{Mantz:2017a}
{Mantz}, A.~B., {Abdulla}, Z., {Allen}, S.~W., {et~al.} 2017, ArXiv e-prints,
  XXL Paper XVII

\bibitem[{{Mantz} {et~al.}(2014){Mantz}, {Abdulla}, {Carlstrom}, {Greer},
  {Leitch}, {Marrone}, {Muchovej}, {Adami}, {Birkinshaw}, {Bremer}, {Clerc},
  {Giles}, {Horellou}, {Maughan}, {Pacaud}, {Pierre}, \&
  {Willis}}]{Mantz:2014a}
{Mantz}, A.~B., {Abdulla}, Z., {Carlstrom}, J.~E., {et~al.} 2014, \apj, 794,
  157, XXL Paper V

\bibitem[{{Mart{\'{\i}}nez} {et~al.}(2010){Mart{\'{\i}}nez}, {Del Olmo},
  {Coziol}, {Perea}, \& {Focardi}}]{Martinez:2010a}
{Mart{\'{\i}}nez}, M., {Del Olmo}, A., {Coziol}, R., {Perea}, J., \& {Focardi},
  P. 2010, in Astronomical Society of the Pacific Conference Series, Vol. 421,
  Galaxies in Isolation: Exploring Nature Versus Nurture, ed.
  L.~{Verdes-Montenegro}, A.~{Del Olmo}, \& J.~{Sulentic}, 125

\bibitem[{{Martini} {et~al.}(2013){Martini}, {Miller}, {Brodwin}, {Stanford},
  {Gonzalez}, {Bautz}, {Hickox}, {Stern}, {Eisenhardt}, {Galametz}, {Norman},
  {Jannuzi}, {Dey}, {Murray}, {Jones}, \& {Brown}}]{Martini:2013a}
{Martini}, P., {Miller}, E.~D., {Brodwin}, M., {et~al.} 2013, \apj, 768, 1

\bibitem[{{Marziani} {et~al.}(2017){Marziani}, {D'Onofrio}, {Bettoni},
  {Poggianti}, {Moretti}, {Fasano}, {Fritz}, {Cava}, {Varela}, \&
  {Omizzolo}}]{Marziani:2017a}
{Marziani}, P., {D'Onofrio}, M., {Bettoni}, D., {et~al.} 2017, \aap, 599, A83

\bibitem[{{Maughan} \& {Reiprich}(2018)}]{Maughan:2018a}
{Maughan}, B. \& {Reiprich}, T.~H. 2018, In prep

\bibitem[{{Maughan} {et~al.}(2012){Maughan}, {Giles}, {Randall}, {Jones}, \&
  {Forman}}]{Maughan:2012a}
{Maughan}, B.~J., {Giles}, P.~A., {Randall}, S.~W., {Jones}, C., \& {Forman},
  W.~R. 2012, \mnras, 421, 1583

\bibitem[{{Mehrtens} {et~al.}(2012){Mehrtens}, {Romer}, {Hilton},
  {Lloyd-Davies}, {Miller}, {Stanford}, {Hosmer}, {Hoyle}, {Collins}, {Liddle},
  {Viana}, {Nichol}, {Stott}, {Dubois}, {Kay}, {Sahl{\'e}n}, {Young}, {Short},
  {Christodoulou}, {Watson}, {Davidson}, {Harrison}, {Baruah}, {Smith},
  {Burke}, {Mayers}, {Deadman}, {Rooney}, {Edmondson}, {West}, {Campbell},
  {Edge}, {Mann}, {Sabirli}, {Wake}, {Benoist}, {da Costa}, {Maia}, \&
  {Ogando}}]{Mehrtens:2012a}
{Mehrtens}, N., {Romer}, A.~K., {Hilton}, M., {et~al.} 2012, \mnras, 423, 1024

\bibitem[{{Merloni} {et~al.}(2012){Merloni}, {Predehl}, {Becker},
  {B{\"o}hringer}, {Boller}, {Brunner}, {Brusa}, {Dennerl}, {Freyberg},
  {Friedrich}, {Georgakakis}, {Haberl}, {Hasinger}, {Meidinger}, {Mohr},
  {Nandra}, {Rau}, {Reiprich}, {Robrade}, {Salvato}, {Santangelo}, {Sasaki},
  {Schwope}, {Wilms}, \& {German eROSITA Consortium}}]{Merloni:2012a}
{Merloni}, A., {Predehl}, P., {Becker}, W., {et~al.} 2012, ArXiv e-prints

\bibitem[{{Nastasi} {et~al.}(2011){Nastasi}, {Fassbender}, {B{\"o}hringer},
  {{\v S}uhada}, {Rosati}, {Pierini}, {Verdugo}, {Santos}, {Schwope}, {de
  Hoon}, {Kohnert}, {Lamer}, {M{\"u}hlegger}, \& {Quintana}}]{Nastasi:2011a}
{Nastasi}, A., {Fassbender}, R., {B{\"o}hringer}, H., {et~al.} 2011, \aap, 532,
  L6

\bibitem[{{Pacaud} {et~al.}(2016){Pacaud}, {Clerc}, {Giles}, {Adami},
  {Sadibekova}, {Pierre}, {Maughan}, {Lieu}, {Le F{\`e}vre}, {Alis}, {Altieri},
  {Ardila}, {Baldry}, {Benoist}, {Birkinshaw}, {Chiappetti},
  {D{\'e}mocl{\`e}s}, {Eckert}, {Evrard}, {Faccioli}, {Gastaldello}, {Guennou},
  {Horellou}, {Iovino}, {Koulouridis}, {Le Brun}, {Lidman}, {Liske},
  {Maurogordato}, {Menanteau}, {Owers}, {Poggianti}, {Pomar{\`e}de}, {Pompei},
  {Ponman}, {Rapetti}, {Reiprich}, {Smith}, {Tuffs}, {Valageas}, {Valtchanov},
  {Willis}, \& {Ziparo}}]{Pacaud:2016a}
{Pacaud}, F., {Clerc}, N., {Giles}, P.~A., {et~al.} 2016, \aap, 592, A2, XXL
  Paper II

\bibitem[{{Pacaud} {et~al.}(2007){Pacaud}, {Pierre}, {Adami}, {Altieri},
  {Andreon}, {Chiappetti}, {Detal}, {Duc}, {Galaz}, {Gueguen}, {Le F{\`e}vre},
  {Hertling}, {Libbrecht}, {Melin}, {Ponman}, {Quintana}, {Refregier},
  {Sprimont}, {Surdej}, {Valtchanov}, {Willis}, {Alloin}, {Birkinshaw},
  {Bremer}, {Garcet}, {Jean}, {Jones}, {Le F{\`e}vre}, {Maccagni}, {Mazure},
  {Proust}, {R{\"o}ttgering}, \& {Trinchieri}}]{Pacaud:2007a}
{Pacaud}, F., {Pierre}, M., {Adami}, C., {et~al.} 2007, \mnras, 382, 1289

\bibitem[{{Pacaud} {et~al.}(2006){Pacaud}, {Pierre}, {Refregier}, {Gueguen},
  {Starck}, {Valtchanov}, {Read}, {Altieri}, {Chiappetti}, {Gandhi}, {Garcet},
  {Gosset}, {Ponman}, \& {Surdej}}]{Pacaud:2006a}
{Pacaud}, F., {Pierre}, M., {Refregier}, A., {et~al.} 2006, \mnras, 372, 578

\bibitem[{{Pierre} {et~al.}(2016){Pierre}, {Pacaud}, {Adami}, {Alis},
  {Altieri}, {Baran}, {Benoist}, {Birkinshaw}, {Bongiorno}, {Bremer}, {Brusa},
  {Butler}, {Ciliegi}, {Chiappetti}, {Clerc}, {Corasaniti}, {Coupon}, {De
  Breuck}, {Democles}, {Desai}, {Delhaize}, {Devriendt}, {Dubois}, {Eckert},
  {Elyiv}, {Ettori}, {Evrard}, {Faccioli}, {Farahi}, {Ferrari}, {Finet},
  {Fotopoulou}, {Fourmanoit}, {Gandhi}, {Gastaldello}, {Gastaud},
  {Georgantopoulos}, {Giles}, {Guennou}, {Guglielmo}, {Horellou}, {Husband},
  {Huynh}, {Iovino}, {Kilbinger}, {Koulouridis}, {Lavoie}, {Le Brun}, {Le
  Fevre}, {Lidman}, {Lieu}, {Lin}, {Mantz}, {Maughan}, {Maurogordato},
  {McCarthy}, {McGee}, {Melin}, {Melnyk}, {Menanteau}, {Novak}, {Paltani},
  {Plionis}, {Poggianti}, {Pomarede}, {Pompei}, {Ponman}, {Ramos-Ceja},
  {Ranalli}, {Rapetti}, {Raychaudury}, {Reiprich}, {Rottgering}, {Rozo},
  {Rykoff}, {Sadibekova}, {Santos}, {Sauvageot}, {Schimd}, {Sereno}, {Smith},
  {Smol{\v c}i{\'c}}, {Snowden}, {Spergel}, {Stanford}, {Surdej}, {Valageas},
  {Valotti}, {Valtchanov}, {Vignali}, {Willis}, \& {Ziparo}}]{Pierre:2016a}
{Pierre}, M., {Pacaud}, F., {Adami}, C., {et~al.} 2016, \aap, 592, A1, XXL
  Paper I

\bibitem[{{Pierre} {et~al.}(2006){Pierre}, {Pacaud}, {Duc}, {Willis},
  {Andreon}, {Valtchanov}, {Altieri}, {Galaz}, {Gueguen}, {Le F{\`e}vre},
  {F{\`e}vre}, {Ponman}, {Sprimont}, {Surdej}, {Adami}, {Alshino}, {Bremer},
  {Chiappetti}, {Detal}, {Garcet}, {Gosset}, {Jean}, {Maccagni}, {Marinoni},
  {Mazure}, {Quintana}, \& {Read}}]{Pierre:2006a}
{Pierre}, M., {Pacaud}, F., {Duc}, P.-A., {et~al.} 2006, \mnras, 372, 591

\bibitem[{{Pierre} {et~al.}(2004){Pierre}, {Valtchanov}, {Altieri}, {Andreon},
  {Bolzonella}, {Bremer}, {Disseau}, {Dos Santos}, {Gandhi}, {Jean}, {Pacaud},
  {Read}, {Refregier}, {Willis}, {Adami}, {Alloin}, {Birkinshaw}, {Chiappetti},
  {Cohen}, {Detal}, {Duc}, {Gosset}, {Hjorth}, {Jones}, {Le F{\`e}vre},
  {Lonsdale}, {Maccagni}, {Mazure}, {McBreen}, {McCracken}, {Mellier},
  {Ponman}, {Quintana}, {Rottgering}, {Smette}, {Surdej}, {Starck}, {Vigroux},
  \& {White}}]{Pierre:2004a}
{Pierre}, M., {Valtchanov}, I., {Altieri}, B., {et~al.} 2004, {\jcap}, 9, 011

\bibitem[{{Pillepich} {et~al.}(2012){Pillepich}, {Porciani}, \&
  {Reiprich}}]{Pillepich:2012a}
{Pillepich}, A., {Porciani}, C., \& {Reiprich}, T.~H. 2012, \mnras, 422, 44

\bibitem[{{Planck Collaboration} {et~al.}(2014){Planck Collaboration}, {Ade},
  {Aghanim}, {Armitage-Caplan}, {Arnaud}, {Ashdown}, {Atrio-Barandela},
  {Aumont}, {Baccigalupi}, {Banday}, \& et~al.}]{Planck-Collaboration:2014a}
{Planck Collaboration}, {Ade}, P.~A.~R., {Aghanim}, N., {et~al.} 2014, \aap,
  571, A20

\bibitem[{{Pratt} {et~al.}(2009){Pratt}, {Croston}, {Arnaud}, \&
  {B{\"o}hringer}}]{Pratt:2009a}
{Pratt}, G.~W., {Croston}, J.~H., {Arnaud}, M., \& {B{\"o}hringer}, H. 2009,
  \aap, 498, 361

\bibitem[{{Rosati} {et~al.}(1998){Rosati}, {Della Ceca}, {Norman}, \&
  {Giacconi}}]{Rosati:1998a}
{Rosati}, P., {Della Ceca}, R., {Norman}, C., \& {Giacconi}, R. 1998, {\apjl},
  492, L21

\bibitem[{{Rozo} {et~al.}(2010){Rozo}, {Wechsler}, {Rykoff}, {Annis}, {Becker},
  {Evrard}, {Frieman}, {Hansen}, {Hao}, {Johnston}, {Koester}, {McKay},
  {Sheldon}, \& {Weinberg}}]{Rozo:2010a}
{Rozo}, E., {Wechsler}, R.~H., {Rykoff}, E.~S., {et~al.} 2010, \apj, 708, 645

\bibitem[{{Santos} {et~al.}(2011){Santos}, {Fassbender}, {Nastasi},
  {B{\"o}hringer}, {Rosati}, {{\v S}uhada}, {Pierini}, {Nonino},
  {M{\"u}hlegger}, {Quintana}, {Schwope}, {Lamer}, {de Hoon}, \&
  {Strazzullo}}]{Santos:2011a}
{Santos}, J.~S., {Fassbender}, R., {Nastasi}, A., {et~al.} 2011, \aap, 531, L15

\bibitem[{{Sehgal} {et~al.}(2013){Sehgal}, {Addison}, {Battaglia},
  {Battistelli}, {Bond}, {Das}, {Devlin}, {Dunkley}, {D{\"u}nner}, {Gralla},
  {Hajian}, {Halpern}, {Hasselfield}, {Hilton}, {Hincks}, {Hlozek}, {Hughes},
  {Kosowsky}, {Lin}, {Louis}, {Marriage}, {Marsden}, {Menanteau}, {Moodley},
  {Niemack}, {Page}, {Partridge}, {Reese}, {Sherwin}, {Sievers}, {Sif{\'o}n},
  {Spergel}, {Staggs}, {Swetz}, {Switzer}, \& {Wollack}}]{Sehgal:2013a}
{Sehgal}, N., {Addison}, G., {Battaglia}, N., {et~al.} 2013, \apj, 767, 38

\bibitem[{{Smith} {et~al.}(2001){Smith}, {Brickhouse}, {Liedahl}, \&
  {Raymond}}]{Smith:2001a}
{Smith}, R.~K., {Brickhouse}, N.~S., {Liedahl}, D.~A., \& {Raymond}, J.~C.
  2001, \apjl, 556, L91

\bibitem[{{Smol{\v c}i{\'c}} {et~al.}(2016){Smol{\v c}i{\'c}}, {Delhaize},
  {Huynh}, {Bondi}, {Ciliegi}, {Novak}, {Baran}, {Birkinshaw}, {Bremer},
  {Chiappetti}, {Ferrari}, {Fotopoulou}, {Horellou}, {McGee}, {Pacaud},
  {Pierre}, {Raychaudhury}, {R{\"o}ttgering}, \& {Vignali}}]{Smolcic:2016a}
{Smol{\v c}i{\'c}}, V., {Delhaize}, J., {Huynh}, M., {et~al.} 2016, \aap, 592,
  A10, XXL Paper XI

\bibitem[{{Stanford} {et~al.}(2014){Stanford}, {Gonzalez}, {Brodwin},
  {Gettings}, {Eisenhardt}, {Stern}, \& {Wylezalek}}]{Stanford:2014a}
{Stanford}, S.~A., {Gonzalez}, A.~H., {Brodwin}, M., {et~al.} 2014, \apjs, 213,
  25

\bibitem[{{Tasse} {et~al.}(2008){Tasse}, {Le Borgne}, {R{\"o}ttgering}, {Best},
  {Pierre}, \& {Rocca-Volmerange}}]{Tasse:2008a}
{Tasse}, C., {Le Borgne}, D., {R{\"o}ttgering}, H., {et~al.} 2008, \aap, 490,
  879

\bibitem[{{Taylor}(2005)}]{Taylor:2005a}
{Taylor}, M.~B. 2005, in Astronomical Society of the Pacific Conference Series,
  Vol. 347, Astronomical Data Analysis Software and Systems XIV, ed.
  P.~{Shopbell}, M.~{Britton}, \& R.~{Ebert}, 29

\bibitem[{{Willis} {et~al.}(2013){Willis}, {Clerc}, {Bremer}, {Pierre},
  {Adami}, {Ilbert}, {Maughan}, {Maurogordato}, {Pacaud}, {Valtchanov},
  {Chiappetti}, {Thanjavur}, {Gwyn}, {Stanway}, \& {Winkworth}}]{Willis:2013a}
{Willis}, J.~P., {Clerc}, N., {Bremer}, M.~N., {et~al.} 2013, \mnras, 430, 134

\end{thebibliography}
